\documentclass[eqsecnum,floats,twocolumn, english,prd,aps,nofootinbib,superscriptaddress,amsfonts,longbibliography,tightenlines,11pt]{revtex4-1}
\usepackage[T1]{fontenc}
\usepackage[latin9]{inputenc}
\usepackage{color}
\usepackage{babel}
\usepackage{mathrsfs}
\usepackage{amsmath}
\usepackage{amssymb}
\usepackage{graphicx}
\usepackage{esint}
\usepackage[unicode=true,pdfusetitle,bookmarks=true,bookmarksnumbered=false,bookmarksopen=false, breaklinks=false,pdfborder={0 0 1},backref=false,colorlinks=false] {hyperref}

\usepackage{breakurl}

\makeatletter

\allowdisplaybreaks[1]

\makeatother

\begin{document}

\title{Callan-Giddings-Harvey-Strominger vacuum in loop quantum gravity and singularity resolution}

\author{Alejandro Corichi}

\email{corichi@matmor.unam.mx}

\affiliation{Centro de Ciencias Matemtic\'{a}s, Universidad Nacional Aut\'{o}noma de M\'{e}xico, Campus Morelia, Apartado Postal 61-3, Morelia, Michoac\'{a}n 58090, Mexico}
\affiliation{Center for Fundamental Theory, Institute for Gravitation and the Cosmos, Pennsylvania State University, University Park, PA 16802, USA}

\author{Javier Olmedo}

\email{jolmedo@lsu.edu}

\affiliation{Department of Physics and Astronomy, Louisiana State University, Baton Rouge, Louisiana 70803-4001, USA}
\affiliation{Instituto de F\'{i}sica, Facultad de Ciencias, Igu\'{a} 4225, Montevideo, Uruguay}

\author{Saeed Rastgoo}

\email{saeed@xanum.uam.mx}

\affiliation{Departamento de F\'{i}sica, Universidad Aut\'{o}noma Metropolitana - Iztapalapa\\ San Rafael Atlixco 186, M\'{e}xico D.F. 09340, M\'{e}xico}
\affiliation{Centro de Ciencias Matemtic\'{a}s, Universidad Nacional Aut\'{o}noma de M\'{e}xico, Campus Morelia, Apartado Postal 61-3, Morelia, Michoac\'{a}n 58090, Mexico}

\date{\today}

\begin{abstract}
        We study here a complete quantization of a Callan-Giddings-Harvey-Strominger
        (CGHS) vacuum model following loop quantum gravity techniques. Concretely,
        we adopt a formulation of the model in terms of a set of new variables
        that resemble the ones commonly employed in spherically symmetric
        loop quantum gravity. The classical theory consists of two pairs of
        canonical variables plus a scalar and diffeomorphism (first class)
        constraints. We consider a suitable redefinition of the Hamiltonian
        constraint such that the new constraint algebra (with structure constants)
        is well adapted to the Dirac quantization approach. For it, we adopt
        a polymeric representation for both the geometry and the dilaton field.
        On the one hand, we find a suitable invariant domain of the scalar
        constraint operator, and we construct explicitly its solution space.
        There, the eigenvalues of the dilaton and the metric operators cannot
        vanish locally, allowing us to conclude that singular geometries are
        ruled out in the quantum theory. On the other hand, the physical Hilbert
        space is constructed out of them, after group averaging the previous
        states with the diffeomorphism constraint. In turn, we identify
        the standard observable corresponding to the mass of the black hole
        at the boundary, in agreement with the classical theory. 
        We also construct an additional
        observable on the bulk associated with the square of the dilaton field,
        with no direct classical analog.
\end{abstract}

\maketitle

\section{Introduction\label{sec:Introduction}}

Since the early days of the search for a quantum theory of gravity,
there has always been the expectation that one of the results that
such a complete theory will yield, would be the
resolution of the spacetime singularities. The first and simplest
reason for this argument is that singularities are in a way, places
where general relativity and the continuous description of spacetime
break down. As in other instances in the history of physics, this
is the regime where one should look for a new theory. Obviously, any
of those new theories should be able to produce the previously known
results of the old theory and also be able to describe the physics
in the regime where the old theory broke down.

Owing to the fact that working with the full theory is, so far, intractable,
it has become standard practice to work with lower dimensional models
or symmetry reduced ones, since generally this allows more control
over the analysis. One of these systems is the well known Callan-Giddings-Harvey-Strominger
(CGHS) model \citep{C.G.Callan1992}. It is a two dimensional dilatonic
model that, in spite of being simpler and classically solvable, has
nontrivial and interesting properties such as a black hole solution,
Hawking radiation, etc. It has been proven to be a very convenient model
for testing some of the quantum gravity ideas in the past, and it has
been subject to many analyses over the past 20 years \cite{Grumiller2002,Fabbri2005}
which has shed some light on the properties of the quantum theory
of the full 4D theory. In particular, additional studies of
 the classical \cite{Kloesch1996} and the semiclassical \cite{A.Ashtekar2011}
regimes of this model, as well as several studies of its quantization \cite{Louis-Martinez1997,Kuchar1997,Varadarajan1998},
have yielded a deeper understanding of some of the interesting physical phenomena in this
toy model that can be expected to be valid also in more realistic situations,
like 4D black holes. However, there are still several questions that remain unanswered, one of them being the way in which a quantum theory of gravity resolves the classical singularity.

In this article, we study the quantization of the CGHS model in a new perspective, namely within the framework of loop
quantum gravity (LQG) \cite{ash-lew,thiem,lqg-gp}. This programme
pursues a background independent non-perturbative quantization
of gravity. It provides a robust kinematical framework
\cite{lost}, while the dynamics has not been completely implemented.
The application of LQG quantization techniques to simpler models,
known as loop quantum cosmology (LQC), has dealt with the question
of the resolution of the singularity at different levels in models
similar to the one under study ---see for instance Refs. \cite{Modesto2004,Ashtekar2005,Boehmer2007,Corichi2015,Gambini2008,A.Ashtekar2011,Ashtekar2011,Gambini2013a,R.Gambini2013}
among others---. In particular, we will pay special attention to Refs. \cite{Gambini2013a,R.Gambini2013},
where a complete quantization of a 3+1 vacuum
spherically symmetric spacetime has been provided, and the singularity of the
model is resolved in a very specific manner. The concrete mechanisms
are based on the requirement of self-adjointness of some observables
of the model, and on the fact that, at the early stages of the quantization,
there is a natural restriction to a subspace of the kinematical Hilbert
space whose states correspond to eigenstates of the triad operators
with non-vanishing eigenvalues, from which the evolution is completely
determined.

The purpose of the present work is to put forward a quantization of
the CGHS model, by extending the methods of \cite{R.Gambini2013}
to the case at hand. A study of the dilatonic systems in the lines of Poisson sigma models in LQG was already carried out in Ref. \cite{bojo-reyes}. The feasibility of the project we are considering
rests on a classical result that allows to cast the CGHS model in
the so-called polar-type variables \cite{Bojowald2006}, similar
to the ones used for the 3+1 spherically symmetric case. These variables
were introduced in \cite{us-new-var,Rastgoo2013} and were further  generalized in \cite{CGHS-CL-New}. Concretely, one
introduces a triadic description of the model for the geometry, together
with a canonical transformation in order to achieve a description
as similar as possible to the one of Ref. \cite{Bojowald2006} in
3+1 at the kinematical level. Then, after considering some second
class conditions and solving the Gauss constraint classically, one
ends with a totally first class system with a Hamiltonian
and a diffeomorphism constraint. Furthermore, based on a proposal  in Refs. \cite{Rastgoo2013,CGHS-CL-New}, a redefinition of the scalar constraint is made, such that this constraint admits the standard algebra with the diffeomorphism constraint, while having a vanishing  brackets with itself. In this situation,
we can follow similar arguments to those in \cite{R.Gambini2013}
to achieve a complete quantization of the CGHS model, showing that
the quantum theory provides a description where the singularity is
resolved in a certain way. Additionally some new observable emerge in the quantum regime, which have no classical analogue.

The structure of this paper is as follows: in section \ref{sec:Brief-review},
we present a very brief review of the CGHS model to show that it contains
a black hole solution with a singularity. Section \ref{sec:Similarity}
is dedicated to recall how one can derive polar-type variables
for the Hamiltonian formulation of the CGHS and 3+1 spherically symmetric
models from a generic 2D dilatonic action, and thus showing the underlying
similarity between the two models in these variables. In section \ref{sec:Preparing},
we illustrate a way to turn the Dirac algebra of the constraint in
the CGHS into a Lie algebra, and thus preparing it for the Dirac quantization.
Section \ref{sec:Quantization} is about quantization: we first introduce
the kinematical Hilbert space of the theory in \ref{sub:kinematical-Hilbert},
we then represent the Hamiltonian constraint on this space in \ref{sub:Representation},
and in subsection \ref{sub:Sing-resolv}, we argue about the resolution
of the singularity in CGHS. Then, we put forward a discussion about the properties
of the solutions to the Hamiltonian constraint in section \ref{sub:properties-sol}.
Finally we note in section \ref{sub:observables} that the same observables
first derived in \cite{Gambini2013a} can also be introduced here.

\section{Brief review of the CGHS model\label{sec:Brief-review}}

The CGHS model \citep{C.G.Callan1992} is a 2D dilatonic model. It has a black hole solution, Hawking radiation, and is classically solvable.
This, together with the fact that it is easier to handle than the full
4D theory or many other models, makes it a powerful test-bench for
many of the ideas in quantum gravity. There has been an extensive previous
work on this model in the literature in the classical and the quantum/semiclassical
regime.

\begin{figure*}
                \centering
        \includegraphics[scale=0.5]{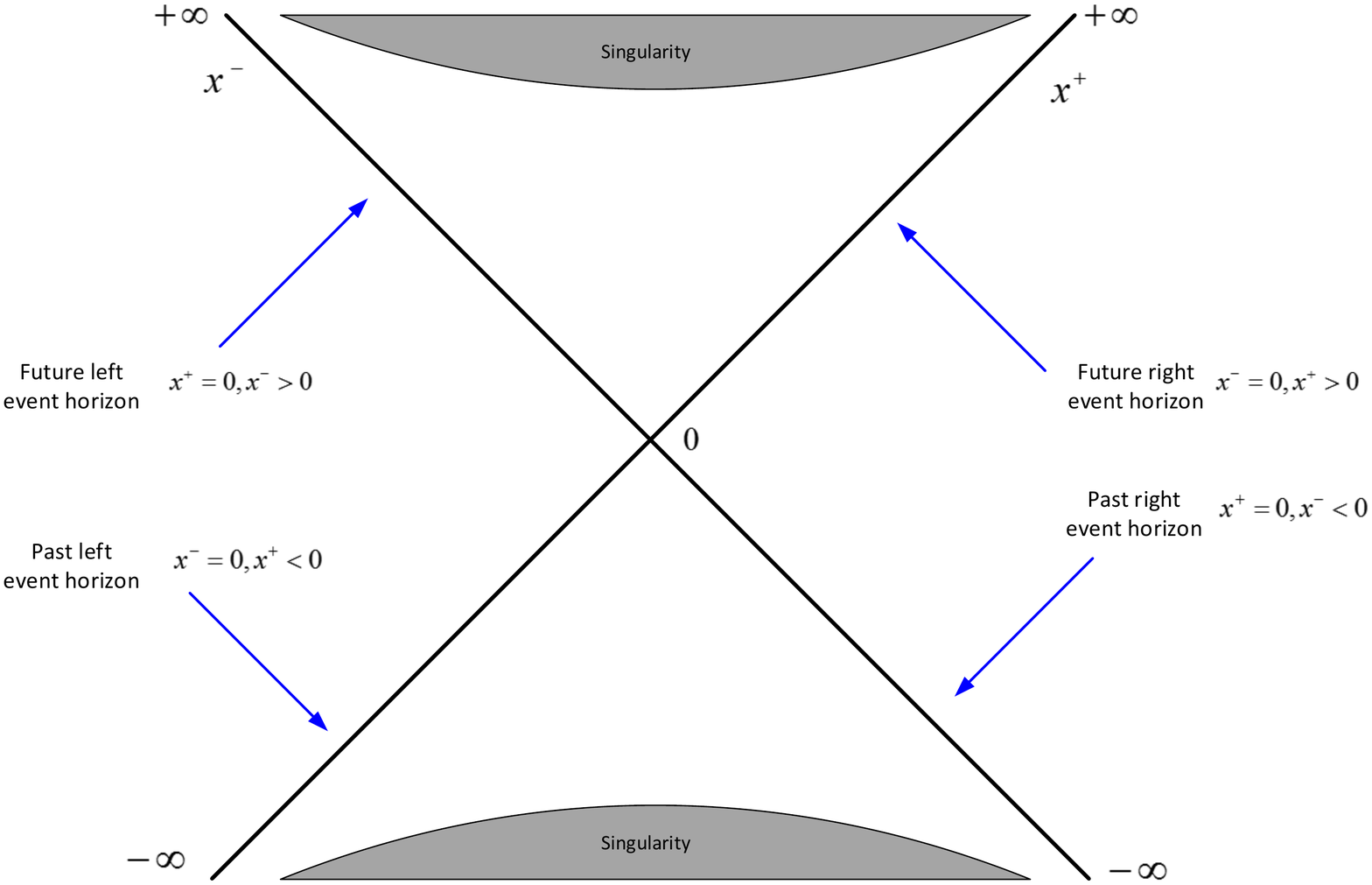}
        \caption{ The Kruskal diagram of the CGHS black hole without matter field\label{fig:Kruskal-CGHS}}
\end{figure*}

The CGHS action is 
\begin{align}\nonumber
        S_{\textrm{g-CGHS}}=&\frac{1}{2G_2}\int d^{2}x\sqrt{-|g|}e^{-2\phi}\\
&\times\left(R+4g^{ab}\partial_{a}\phi\partial_{b}\phi+4\lambda^{2}\right),\label{eq:CGHS-act}
\end{align}
where $G_2$ is the 2-dimensional Newton constant, $\phi$ is the dilaton field and $\lambda$ the cosmological constant. In double null coordinates $x^{\pm}=x^{0}\pm x^{1}$ and
in conformal gauge 
\begin{align}
        g_{+-}=-\frac{1}{2}e^{2\rho} & , & g_{--}=g_{++}=0,
\end{align}
the solution is 
\begin{equation}
        e^{-2\rho}=e^{-2\phi}=\frac{G_2M}{\lambda}-\lambda^{2}x^{+}x^{-},
\end{equation}
where $M$ is a constant of integration which can be identified as
the ADM (at spatial infinity) or the Bondi (at null infinity) mass.
The scalar curvature turns out to be 
\begin{equation}
        R=\frac{4G_2M\lambda}{\frac{G_2M}{\lambda}-\lambda^{2}x^{+}x^{-}},
\end{equation}
which corresponds a black hole with mass $M$ with a singularity at
\begin{equation}
        x^{+}x^{-}=\frac{G_2M}{\lambda^{3}}.
\end{equation}
The Kruskal diagram of the CGHS black hole is very similar to the
4D Schwarzschild model and is depicted in figure \ref{fig:Kruskal-CGHS}.

\section{Similarity of the CGHS and 3+1 spherically symmetric models\label{sec:Similarity}}

As we mentioned in the section \ref{sec:Introduction}, the key point
of the ability to extend the results of \citep{R.Gambini2013} to
the CGHS model is writing the latter in polar-type variables
\citep{Bojowald2006}. This has been mainly done in \citep{us-new-var,Rastgoo2013,CGHS-CL-New}.
Here, we give a brief review of this formulation and its key similarities and differences to the 3+1 spherically-symmetric gravity.

Let us start by considering the most generic 2D diffeomorphism-invariant action yielding second order
differential equations for the metric $g$ and a scalar (dilaton)
field $\Phi$ \citep{Grumiller2002} 
\begin{align}\nonumber
        S=&\frac{1}{G_2}\int d^{2}x\sqrt{-|g|}\\
&\times\left(Y(\Phi)R(g)+V\left(\left(\nabla\Phi\right)^{2},\Phi\right)\right).
\end{align}
Within this class we choose a subclass \citep{Banks1991,Odintsov1991,Kloesch1996}
that is generic enough for our purposes, 
\begin{align}\nonumber
        S_{g\textnormal{-dil}}=&\frac{1}{G_2}\int d^{2}x\sqrt{-|g|}\\
        &\times\left(Y(\Phi)R(g)+\frac{1}{2}g^{ab}\partial_{a}\Phi\partial_{b}\Phi+V(\Phi)\right).\label{eq:g-dil-gen}
\end{align}
Here, $Y(\Phi)$ is the non-minimal coupling coefficient, $V(\Phi)$ is the potential of the dilaton field, and $\frac{1}{2}g^{ab}\partial_{a}\Phi\partial_{b}\Phi$ is its kinetic term.
The latter can be removed at will by a conformal transformation.
With the choice $Y(\Phi)=\frac{1}{8}\Phi^{2}$ and $V(\Phi)=\frac{1}{2}\Phi^{2}\lambda^2$, we obtain the CGHS model \citep{C.G.Callan1992}, which is given by the action
\begin{align}\nonumber
        S_{\textrm{CGHS}}= & \frac{1}{G_2}\int d^{2}x\sqrt{-|g}|\\
        &\times\left(\frac{1}{8}\Phi^{2}R+\frac{1}{2}g^{ab}\partial_{a}\Phi\partial_{b}\Phi+\frac{1}{2}\Phi^{2}\lambda^2\right),\label{eq:dil-CGHS}
\end{align}
with $\lambda$ the cosmological constant. It coincides with Eq. (\ref{eq:CGHS-act}) for $\Phi=2e^{-\phi}$. 

In the same way, we may notice the parallelism with 3+1 spherically symmetric gravity in vacuum. By using the spherical symmetry ansatz, 
\begin{align}
        ds^{2}= & g_{\mu\nu}dx^{\mu}dx^{\nu}+\Phi^{2}(d\theta^{2}+\sin^{2}(\theta)d\phi^{2}), \label{eq:sph-ansatz}
\end{align}
with $\mu,\nu=0,1$, for the metric of the 4D model, its action can be written as
\begin{align}\nonumber
 S_{\textrm{spher}}= & \frac{1}{G}\int d^{2}x\sqrt{-|g|}\\
 &\times\left(\frac{1}{4}\Phi^{2}R+\frac{1}{2}g^{ab}\partial_{a}\Phi\partial_{b}\Phi+\frac{1}{2}\right),\label{eq:dil-spher}
\end{align}
where $G$ is the Newton's constant in 4D Einstein's theory. One can see that this is identical to (\ref{eq:g-dil-gen}) if one chooses $Y(\Phi)=\frac{1}{4}\Phi^{2}$,  $V(\Phi)=\frac{1}{2}$, and replaces $G_2$ with $G$. Note that although
 the actions of both CGHS and 4D models contain the variable $\Phi$,  the interpretation
of this variable is different in each of these cases. In 4D spherical
gravity,  $\Phi$ is actually a part of the metric,  the coefficient
multiplied by the two-sphere part of the metric as can be seen from
(\ref{eq:sph-ansatz}). In the CGHS, however, it is a non-geometric
degree of freedom corresponding to the scalar dilaton field.

\subsection{Polar-type variables for spherically symmetric model}

As can be seen in details in \citep{us-new-var}, one can write  (\ref{eq:dil-spher}) in terms of the polar-type
variables. Here we 
only  explain the procedure briefly. In 3+1 spherically symmetric case,
one first removes the dilaton kinetic term by a conformal transformation
and then writes the theory in tetrad variables. One then adds the
torsion free condition, multiplied by a Lagrange multiplier $X^{I}$,
to the Lagrangian. Here $I$ is a Lorentz internal index representing
the internal local gauge group of the theory. One then makes an integration
by parts such that derivatives of $X^{I}$ appear in the Lagrangian.
After ADM decomposition of the action and some further calculations, it turns
out that the configuration variables are 
\begin{align}
        \left\{ ^{*}X^{I}=\epsilon^{IJ}X_{J},\omega_{1}\right\},  &  & I,J=\{0,1\}
\end{align}
where $\omega_{1}$ is the  spatial part of the spin connection. The
corresponding momenta then will be 
\begin{align}
        P_{I}= & \frac{\partial\mathcal{L}}{\partial{}^{*}\dot{X}^{I}}=2\sqrt{q}n_{I},\label{eq:PI}\\
        P_{\omega}= & \frac{\partial\mathcal{L}}{\partial\dot{\omega}_{1}}=\frac{1}{2}\Phi^{2}.
\end{align}
Here $n_{I}=n_{\mu}e^{\mu}{}_{I}$ is the $I$'th (internal) component
of the normal to the spatial hypersurface, with $e^{\mu}{}_{I}$ being
the  tetrad, and $q$ is the determinant of the spatial metric.
Then by a Legendre transformation one can arrive at the Hamiltonian
in these variables. From this Hamiltonian one can get to the Hamiltonian
in polar-type variables by considering the following relation
\begin{equation}
        \Vert P\Vert^{2}=-|P|^{2}=-\eta^{IJ}P_{I}P_{J}=
        4q
        .\label{eq:qPE-rel-sph}
\end{equation}
Then we adopt the parametrization
\begin{equation}
        q=\frac{\left(E^{\varphi}\right)^{2}}{(E^{x})^{\frac{1}{2}}},
\end{equation}
based on the form of the 3+1 metric in terms of the polar-type variables. Equation (\ref{eq:qPE-rel-sph}) leads to the following canonical
transformation to polar-type variables 
\begin{align}
        P_{\omega}= & E^{x},\\
        \Vert P\Vert= & \frac{2E^{\varphi}}{(E^{x})^{\frac{1}{4}}},\\
        P_{0}= & \frac{2E^{\varphi}}{(E^{x})^{\frac{1}{4}}}\cosh(\eta),\\
        P_{1}= & \frac{2E^{\varphi}}{(E^{x})^{\frac{1}{4}}}\sinh(\eta).
\end{align}
The first equation above is just a renaming, and the rest of them follow
naturally from (\ref{eq:qPE-rel-sph}). By finding a generating function
for this canonical transformation, one can find the corresponding
canonical variables $\{K_{x},K_{\varphi},Q_{\eta}\}$ to the above
momenta $\{E^{x},E^{\varphi},\eta\}$ and then write the Hamiltonian in
these variables. The Hamiltonian will be the sum of three constraints
as expected 
\begin{equation}
        H=\frac{1}{G}\int dx\left(N\mathcal{H}+N^{1}\mathcal{D}+\omega_{0}\mathcal{G}\right)
\end{equation}
where $N$ and $N^{1}$ are lapse and shift respectively, $\omega_{0}$
is the ``time component'' of the spin connection which is another Lagrange
multiplier, and $\mathcal{H},\,\mathcal{D}$ and $\mathcal{G}$ are
Hamiltonian, diffeomorphism and Gauss constraints respectively. In
order to make things simpler, Gambini et. al. in \citep{R.Gambini2013}
take $\eta=1$ and since this is second class with the Gauss constraint,
they can be solved to yield the final Hamiltonian 
\begin{widetext}
\begin{align}
        H=\frac{1}{G}\int dx & \left[N\left(\frac{\left((E^{x})'\right)^{2}}{8\sqrt{E^{x}}E^{\varphi}}-\frac{E^{\varphi}}{2\sqrt{E^{x}}}-2K_{\varphi}\sqrt{E^{x}}K_{x}-\frac{E^{\varphi}K_{\varphi}^{2}}{2\sqrt{E^{x}}}-\frac{\sqrt{E^{x}}(E^{x})'(E^{\varphi})'}{2(E^{\varphi})^{2}}+\frac{\sqrt{E^{x}}(E^{x})''}{2E^{\varphi}}\right)\right.\nonumber \\
        & \left.+N^{1}\left(E^{\varphi}K_{\varphi}'-(E^{x})'K_{x}\right)\vphantom{\frac{K_{\varphi}^{2}}{\sqrt{E^{x}}}}\right].
\end{align}
\end{widetext}
\global\long\def\sgn{\operatorname{sgn}}
\global\long\def\Int{\operatorname{Int}}

\subsection{Polar-type variables for the CGHS model}

By guidance from the procedure done in the spherically symmetric case,
one can arrive at similar variables for the CGHS model. Most of the steps are
in principle similar, but there are also some important differences.
The details can be found in \citep{Rastgoo2013} and, again, we will describe
the process in a brief manner. First, we should mention that, although
almost all the studies of the CGHS model have utilized a conformal
transformation to remove the dilaton kinetic term in an effort to render
the theory as a first class system, we will proceeded instead with that
term present. The main reason was that, in this way, the variables will admit a
natural geometrical interpretation, and the quantization of the model can be 
carried out following the ideas of loop quantum gravity. The geometric implications
can be read more easily and directly. In any case, this is just
a choice and it is not of crucial importance.

It turns out that, by following the same procedure of adding the torsion
free condition, writing in tetrad variables and adopting an ADM decomposition, and because the kinetic term (and hence the time derivative)
of the dilaton is present, the configuration variables will be 
\begin{align}
        \left\{ ^{*}X^{I},\omega_{1},\Phi\right\}  &  & I,J=\{0,1\}
\end{align}
with the corresponding momenta 
\begin{align}
        P_{I}= & \frac{\partial\mathcal{L}}{\partial{}^{*}\dot{X}^{I}}=2\sqrt{q}n_{I},\\
        P_{\omega}= & \frac{\partial\mathcal{L}}{\partial\dot{\omega}_{1}}=\frac{1}{4}\Phi^{2}\label{eq:Pomega-CGHS}\\
        P_{\Phi}= & \frac{\partial\mathcal{L}}{\partial\dot{\Phi}}=\frac{\sqrt{q}}{N}\left(N^{1}\Phi'-\dot{\Phi}\right),
\end{align}
where again $N$ and $N^{1}$ are lapse and shift respectively. An important
consequence of these are that (\ref{eq:Pomega-CGHS}) is now a new
primary constraint 
\begin{equation}
        \mu=P_{\omega}-\frac{1}{4}\Phi^{2}\approx0.\label{eq:mu-old}
\end{equation}
In the next step, by making a Legendre transformation, we will get to a
Hamiltonian which now should also contain the new primary constraint
(\ref{eq:mu-old}). To obtain the polar-type variables we
use a similar relation to (\ref{eq:qPE-rel-sph}) which, in the case
of the CGHS model, reads 
\begin{equation}
        \Vert P\Vert^{2}=4q=4\left(E^{\varphi}\right)^{2}\label{eq:qPE-rel-CGHS}
\end{equation}
where we have used again a natural parametrization for $q$ in terms of $E^{\varphi}$ for the CGHS model. Then, we get the new variables 
\begin{align}
        P_{\omega}= & E^{x},\label{eq:PomegaCGHS}\\
        \Vert P\Vert= & 2E^{\varphi},\\
        P_{0}= & 2\cosh(\eta)E^{\varphi},\\
        P_{1}= & 2\sinh(\eta)E^{\varphi}.
\end{align}
Again, they follow naturally from (\ref{eq:qPE-rel-CGHS}) with
a bit of educated guess. These transformations do not affect the pair
$\{\Phi,P_{\Phi}\}$. Once again, by finding a generating function
for this canonical transformation, we can find the corresponding conjugate
variables $\{K_{x},K_{\varphi},Q_{\eta},\Phi\}$ to the above momenta
$\{E^{x},E^{\varphi},\eta,P_{\Phi}\}$ and then write the Hamiltonian in
these variables. The Hamiltonian will be the sum of four constraints
\begin{equation}
        H=\frac{1}{G_2}\int dx\left(N\mathcal{H}+N^{1}\mathcal{D}+\omega_{0}\mathcal{G}+B\mu\right)
\end{equation}
with $B$ being another Lagrange multiplier. Note that, in this case,
unlike the spherically symmetric case, we have 
\begin{equation}
        K_{x}=\omega_{1}.
\end{equation}
Also note that there is an important difference here between the 3+1
spherically-symmetric case and the CGHS model: as a consequence of
what we also mentioned in the beginning of section \ref{sec:Similarity}
and due to (\ref{eq:Pomega-CGHS}) and (\ref{eq:mu-old}), one can
see that $E^{x}$ is classically associated to the
dilaton field in the CGHS model. It has nothing to do with the metric and is a truly distinct
degree of freedom. While, as we mentioned, it is a component of the metric in the 3+1 spherically-symmetric case.

Continuing with the Dirac procedure, since we have a new primary constraint
$\mu$, we need to check its consistency under the evolution. This leads to a new secondary constraint $\alpha$, namely 
\begin{equation}
        \dot{\mu}\approx0\Rightarrow\alpha=K_{\varphi}+\frac{1}{2}\frac{P_{\Phi}\Phi}{E^{\varphi}}\approx0.
\end{equation}
Preservation of $\alpha$ then leads to no new constraint. It turns
out that these two new constraints are second class together 
\begin{equation}
        \{\mu,\alpha\}\not\approx0
\end{equation}
and thus we need to follow the second class Dirac procedure for this case.
So, we solve them to get 
\begin{align}
        \mu=0 & \Rightarrow\Phi=2\sqrt{E^{x}},\\
        \alpha=0 & \Rightarrow P_{\Phi}=-\frac{K_{\varphi}E^{\varphi}}{\sqrt{E^{x}}}.
\end{align}
This eliminates the pair $\{\Phi,P_{\Phi}\}$ in the Hamiltonian.
In order to simplify the process of  quantization, we introduce the new variable
\begin{equation}
        A_{x}=K_{x}-\eta',
\end{equation}
and choose $\eta=1$ which is again second class
with the Gauss constraint. Then, solving these second class constraints together yields an expression for
$Q_{\eta}$ in terms of the remaining variables. In this way, the pair $\{Q_{\eta},\eta\}$ are also eliminated
from the Hamiltonian. A similar procedure has also been done in the
spherically symmetric case in \citep{R.Gambini2013}. Note that we now have
\begin{equation}
        A_{x}=\omega_{1}.
\end{equation}
The Dirac brackets now become
\begin{align}\nonumber
        &\{K_{x}(x),E^{x}(y)\}_{D}  =\{K_{\varphi}(x),E^{\varphi}(y)\}_{D}\\
        &=\{f(x),P_{f}(y)\}_{D}=G_2\delta(x-y),\\
        &\{K_{x}(x),K_{\varphi}(y)\}_{D}  =G_2{\displaystyle \frac{K_{\varphi}}{E^{x}}\delta(x-y)},\\
        &\{K_{x}(x),E^{\varphi}(y)\}_{D} =-G_2{\displaystyle \frac{E^{\varphi}}{E^{x}}\delta(x-y)},
\end{align}
with any other brackets vanishing. These brackets can be brought to  the canonical from
\begin{align}\nonumber
        &\{U_{x}(x),E^{x}(y)\}_{D}=\{K_{\varphi}(x),E^{\varphi}(y)\}_{D}\\
        &=\{f(x),P_{f}(y)\}_{D}=G_2\delta(x-y),
\end{align}
by introducing the redefinition 
\begin{equation}
        U_{x}=K_{x}+\frac{E^{\varphi}K_{\varphi}}{E^{x}}.\label{eq:Ux-Def}
\end{equation}
Finally, we are left with the Hamiltonian
\begin{widetext}
\begin{align}
        H=\frac{1}{G_2}\int dx & \left[N\mathcal{H}+N^{1}\mathcal{D}\right]\nonumber \\
        =\frac{1}{G_2}\int dx & \left[N\left(-K_{\varphi}U_{x}-\frac{E^{\varphi\prime}E^{x\prime}}{E^{\varphi2}}-\frac{1}{2}\frac{E^{x\prime2}}{E^{\varphi}E^{x}}+\frac{E^{x\prime\prime}}{E^{\varphi}}+\frac{1}{2}\frac{K_{\varphi}^{2}E^{\varphi}}{E^{x}}-2E^{\varphi}E^{x}\lambda^{2}\right)\right.           \nonumber\\
          &\left. +N^{1}\left(-U_{x}E^{x\prime}+E^{\varphi}K_{\varphi}'\right)\vphantom{\frac{K_{\varphi}^{2}E^{\varphi}}{E^{x}}}\right].\label{eq:HtotfnUx}
\end{align}
\end{widetext}

\section{Preparing the CGHS Hamiltonian for quantization\label{sec:Preparing}}

At this point, and in order to proceed with the Dirac quantization of the system, we will adopt an Abelianization of the scalar constraint algebra. The reason is the following: the Dirac quantization approach involves several consistency conditions. For instance, the constraint algebra at the quantum level must agree with the classical one. It is well-known that anomalies in the algebra can emerge, and spoil the final quantization. Usually, this situation is more likely to be satisfied  if the constraints fulfill a Lie algebra (with structure constants instead of structure functions of phase-space variables). An even more favorable situations is when (part of the algebra) is strongly Abelian. We already know that the brackets $\{\mathcal{H}(N),\mathcal{D}(N^{1})\}$
and $\{\mathcal{D}(N^{1}),\mathcal{D}(M^{1})\}$  involve structure constants and close under the bracket. But this is not the case for $\{\mathcal{H}(N),\mathcal{H}(M)\}$. Although, in principle, nothing prevents us carry on with the study in this situation, we would like to adopt a strategy based on strong Abelianization that will allow us to complete the quantization, since other choices are either not fully understood or not considerably developed. This strategy consist in a redefinition
of the shift function
\begin{equation}
        \overline{N}^{1}=N^{1}+\frac{NK_{\varphi}}{E^{x\prime}},\label{eq:N1-redef}
\end{equation}
followed by  a redefinition the lapse function as 
\begin{equation}
        \overline{N}=N\frac{E^{\varphi}E^{x}}{E^{x\prime}}.\label{eq:N-redef}
\end{equation}
These yield 
\begin{widetext}
\begin{align}
        H=\frac{1}{G_2}\int dx & \left[\overline{N}\mathcal{H}+\overline{N}^{1}\mathcal{D}\right]\nonumber \\
        =\frac{1}{G_2}\int dx & \;\overline{N}\left[\frac{\partial}{\partial x}\left(\frac{1}{2}\frac{E^{x\prime2}}{E^{\varphi2}E^{x}}-2E^{x}\lambda^{2}-\frac{1}{2}\frac{K_{\varphi}^{2}}{E^{x}}\right)\right]+\overline{N}^{1}\left(-U_{x}E^{x\prime}+E^{\varphi}K_{\varphi}'\right).\label{eq:Htot-deriv}
\end{align}
\end{widetext}
One can check that now
\begin{equation}
        \{\mathcal{H}(N),\mathcal{H}(N^{\prime})\}_{D}=0,
\end{equation}
and thus the Dirac quantization, particularly the loop quantization strategy, is expected to be simpler and potentially successful with respect to other choices considered so far.

We can take advantage of this form of the Hamiltonian constraint and,
by making an integration by parts\footnote{In this work, we ignore the boundary term arising from this integration by parts.}, write Eq. (\ref{eq:Htot-deriv}) as
\begin{align}\nonumber
H=&\frac{1}{G_2}\int dx\;\overline{N}^{\prime}\\\nonumber
&\times\left[\frac{1}{2}\frac{E^{x\prime2}}{E^{\varphi2}E^{x}}-2E^{x}\lambda^{2}-\frac{1}{2}\frac{K_{\varphi}^{2}}{E^{x}}+\lambda  G_2 M\right]\\
&+\overline{N}^{1}\left(-U_{x}E^{x\prime}+E^{\varphi}K_{\varphi}'\right),
\end{align}
where $M$ is the ADM mass of the CGHS black hole and $G_2$ is the dimensionless Newton's constant in 2D spacetimes. 

At this point we are going to first consider the Hamiltonian constraint and prepare it for representation on  the kinematical Hilbert space. Regarding the diffeomorphism constraint, we will adopt the group averaging technique, since, as it is well-known in loop quantum gravity, only finite spatial diffeomorphisms are well-defined unitary operators on the Hilbert space.  

If we rename $\overline{N}^{\prime}\rightarrow N$,
the Hamiltonian constraint can now be written as 
\begin{eqnarray}\nonumber
        \!\!\!\!\!\!\!\!\!&&\mathcal{H}(N)=\frac{1}{G_2}\int dxN\\
        \!\!\!\!\!\!\!\!\!&&\times\left[\frac{1}{2}\frac{E^{x\prime2}}{E^{\varphi2}E^{x}}-2E^{x}\lambda^{2}-\frac{1}{2}\frac{K_{\varphi}^{2}}{E^{x}}+\lambda  G_2 M\right].
\end{eqnarray}
Our final step, before quantization, is to bring the above constraint
in a form that will admit a natural representation on a suitable Hilbert space.
This is achieved by rescaling the lapse function $N\rightarrow2NE^{\varphi}\left(E^{x}\right)^{2}$
such that 
\begin{widetext}
\begin{equation}
        \mathcal{H}(N)=\frac{1}{G_2}\int dxNE^{x}\left[4\left(E^{x}\right)^{2}E^{\varphi}\lambda^{2}+K_{\varphi}^{2}E^{\varphi}-2\lambda  G_2 ME^{\varphi}E^{x}-\frac{\left(E^{x\prime}\right)^{2}}{E^{\varphi}}\right].\label{eq:Hcons-for-quant}
\end{equation}
\end{widetext}

\section{Quantization\label{sec:Quantization}}

\subsection{The kinematical Hilbert space\label{sub:kinematical-Hilbert}}

To quantize the theory, we first need an auxiliary (or kinematical) vector space of states. Then
we should equip it with an inner product and then carry out a Cauchy completion of this space. We will then end up with a kinematical Hilbert space. Afterwards, we need
to find a representation of the phase space variables as operators acting
on this Hilbert space. In order to study the dynamics of the system, since we are dealing with a totally constrained theory, we will follow the Dirac quantization approach. Here, one identifies those quantum structures that are invariant under the gauge symmetries generated by the constraints. In this particular model, we have the group of spatial diffeomorphisms (generated by the diffeomorphism constraint) and the set of time reparametrizations (associated with the Hamiltonian constraint). In the loop representation, only the spatial diffeomorphisms are well understood. Then, we must look for a suitable representation of the Hamiltonian constraint (\ref{eq:Hcons-for-quant}) as a quantum operator,  and look for its kernel which yields a space of  states that  invariant under this constraint. Finally, one should endow this space of solutions with a Hilbert space structure and suitable observables acting on it.

Here, we will adhere to a loop representation for the kinematical variables, except the mass, for which a standard Fock quantization will be adopted. Our full kinematical Hilbert space is the direct  product of two parts, 
\begin{equation}
        \mathscr{H}_{\textrm{kin}}=\mathscr{H}_{\textrm{kin}}^{M}\otimes\left(\bigoplus_{\mathnormal{g}}\mathscr{H}_{\textrm{kin-spin}}^{\mathnormal{g}}\right).
\end{equation}
One part, $\mathscr{H}_{\textrm{kin}}^{M}=L^{2}(\mathbb{R},dM)$,
is associated to the global degree of freedom of the mass of the black
hole $M$. The other part, associated to the gravitational sector, is the direct sum of the spaces, $\mathscr{H}_{\textrm{kin-spin}}^{\mathnormal{g}}$,
each  corresponding to a given
graph (spin network) $\mathnormal{\mathnormal{g}}$ for which we would
like to use the polymer quantization. This choice seems to be natural in 3+1 spherically-symmetric models for the geometrical variables, and due to the parallelism between that model and the CGHS model, we will adopt a similar representation here. 

To construct $\mathscr{H}_{\textrm{kin-spin}}^{\mathnormal{g}}$,
we first take the vector space $Cyl_{\mathnormal{g}}$, of all the
functions of holonomies along the edges of a graph $\mathnormal{g}$,
and the point holonomies ``around'' its vertices, and equip this
vector space with the Haar measure to get the gravitational part of
the kinematical Hilbert space of the given graph $\mathnormal{g}$.
In our case these states are 
\begin{align}\nonumber
        \langle &U_{x},K_{\varphi}|\mathnormal{g},\vec{k},\vec{\mu}\rangle=\prod_{e_{j}\in\mathnormal{g}}\exp\left(\frac{i}{2}k_{j}\int_{e_{j}}dx\,U_{x}(x)\right)\\
&\times\prod_{v_{j}\in\mathnormal{g}}\exp\left(\frac{i}{2}\mu_{j}K_{\varphi}(v_{j})\right).
\end{align}
Here $e_{j}$ are the edges of the graph, $v_{j}$ are its vertices,
$k_{j}\in\mathbb{Z}$ is the edge color, and $\mu_{j}\in\mathbb{R}$
is the vertex color. We indicate the order (i.e. number of the vertices)
of the graph $\mathnormal{g}$ by $V$. Since $\mu_{j}\in\mathbb{R}$,
the above belongs to the space of almost-periodic functions and the
associated Hilbert space will be non-separable.

It is evident that this Hilbert space, $\mathscr{H}_{\textrm{kin-spin}}^{\mathnormal{g}}$,
can be decomposed into a part associated with the normal holonomies
along the edges, which is the space of square summable functions $\ell^{2}$,
and another part associated to the point holonomies, which is the space
of square integrable functions over Bohr-compactified real line with
the associated Haar measure, $L^{2}(\mathbb{R}_{\textrm{Bohr}},d\mu_{\textrm{Haar}})$. The construction for the mass degree of freedom is similar and well-known, and we will not give additional details here. Thus, the full kinematical Hilbert space can be written as 
\begin{align}\nonumber
&\mathscr{H}_{\textrm{kin}}=\mathscr{H}_{\textrm{kin}}^{M}\otimes\left(\bigoplus_{\mathnormal{g}}\mathscr{H}_{\textrm{kin-spin}}^{\mathnormal{g}}\right)=L^{2}(\mathbb{R},dM)\\
&\otimes\left(\bigoplus_{\mathnormal{g}}\left[\bigotimes_{v_j\in g}\ell_j^{2}\otimes L_j^{2}(\mathbb{R}_{\textrm{Bohr}},d\mu_{\textrm{Haar}})\right]\right).
\end{align}
Let us call the kinematical Hilbert space of a single graph $\mathscr{H}_{\textrm{kin}}^{\mathnormal{g}}=\mathscr{H}_{\textrm{kin}}^{M}\otimes\mathscr{H}_{\textrm{kin-spin}}^{\mathnormal{g}}$ (not to be confused with $\mathscr{H}_{\textrm{kin-spin}}^{\mathnormal{g}}$).
There is a basis of states in this Hilbert space denoted by $\{|\mathnormal{g},\vec{k},\vec{\mu},M\rangle\}$.
Then, since now we have a measure, and thus a Hilbert space, we can
define the inner product on $\mathscr{H}_{\textrm{kin}}^{\mathnormal{g}}$
and thus on $\mathscr{H}_{\textrm{kin}}$. As usual in loop quantum
gravity, a spin network defined on $\mathnormal{g}$ can be regarded
as a spin network with support on a larger graph $\bar{\mathnormal{g}}\supset\mathnormal{g}$\textbf{
}by assigning trivial labels to the edges and vertices which are not
in $\mathnormal{g}$. Consequently, for any two graphs $\mathnormal{g}$ and
$\mathnormal{g}^{\prime}$, we take $\bar{\mathnormal{g}}=\mathnormal{g}\cup\mathnormal{g}^{\prime}$
and the inner product of $\mathnormal{g}$ and $\mathnormal{g}^{\prime}$
will be 
\begin{align} &\langle\mathnormal{g},\vec{k},\vec{\mu},M|\mathnormal{g}',\vec{k}',\vec{\mu}',M'\rangle=  \delta(M-M')\\\nonumber
&\times\prod_{\textrm{edges}}\delta_{k_{j},k_{j}^{\prime}}\prod_{\textrm{vertices}}\delta_{\mu_{j},\mu_{j}^{\prime}}
        = \delta(M-M')\delta_{\vec{k},\vec{k}'}\delta_{\vec{\mu},\vec{\mu}'}.
\end{align}
Obviously, the inner product can be extended to arbitrary states by
superposition of the basis states.

\subsection{Representation of operators\label{sub:Representation}}

Now that we have a kinematical Hilbert space, the next step is to
represent the phase space variables on it as operators. We will follow
a similar strategy as the one of Ref. \citep{R.Gambini2013}. First, we choose
the polymerization $K_{\varphi}\to\sin(\rho K_{\varphi})/\rho$. Looking
at (\ref{eq:Hcons-for-quant}), we note that we need to represent
the following phase space variables
\begin{equation}
        E^{x},E^{x\prime},E^{\varphi},\frac{1}{E^{\varphi}},K_{\varphi}^{2}E^{\varphi},M.
\end{equation}
Due to our polymerization scheme and the classical algebra (i.e. Dirac brackets), the momenta can be represented as
\begin{align}
        \widehat{E^{\varphi}}|\mathnormal{g},\vec{k},\vec{\mu},M\rangle=  & \hbar G_2\sum_{v_{j}\in g}\delta(x-x_{j})\mu_{j}|\mathnormal{g},\vec{k},\vec{\mu},M\rangle\label{eq:Ef-rep}\\
        \widehat{E^{x}}|\mathnormal{g},\vec{k},\vec{\mu},M\rangle= & \hbar G_2k_{j}|\mathnormal{g},\vec{k},\vec{\mu},M\rangle,\label{eq:Ex-rep}
\end{align}
where $\hbar G_2$ is the Planck number (recalling that $\hbar$ has dimensions $[LM]$). 
The presence of the Dirac delta function
in (\ref{eq:Ef-rep}) is due to  $E^{\varphi}$ being a density. The global degree of freedom $M$, corresponding to the Dirac observable on the boundary associated
to the mass of the black hole, can
be represented as 
\begin{equation}
        \hat{M}|\mathnormal{g},\vec{k},\vec{\mu},M\rangle=M|\mathnormal{g},\vec{k},\vec{\mu},M\rangle.
\end{equation}
To represent the last contribution in (\ref{eq:Hcons-for-quant}), we combine $E^{x\prime}$ with
$\frac{1}{E^{\varphi}}$, and use the Thiemann's trick \citep{Thiemann1996}, to represent it as 
\begin{align}\nonumber
&\widehat{\left[\frac{[(E^{x})^\prime]^2}{E^{\varphi}}\right]}|g,\vec{k},\vec{\mu},M\rangle=\sum_{v_{j}\in g}\delta(x-x(v_{j}))\\\nonumber
&\times\frac{\sgn(\mu_{j})\hbar G_2}{\rho^{2}}\left(k_{j}-k_{j-1}\right)^2\Big[|\mu_{j}+\rho|^{1/2}\\
&-|\mu_{j}-\rho|^{1/2}\Big]^{2}|g,\vec{k},\vec{\mu},M\rangle.
\end{align}
This is due to the operator $\widehat{N_{n\rho}^{\varphi}}$ corresponding
to $K_{\varphi}$, which is represented by the action of the point
holonomies of length $\rho$ 
\begin{equation}
        \widehat{N_{\pm n\rho}^{\varphi}}(x)|g,\vec{k},\vec{\mu},M\rangle=|g,\vec{k},\vec{\mu}_{\pm n\rho}^{\prime},M\rangle,\quad n\in\mathbb{N}.
\end{equation}
In this expression, the new vector $\vec{\mu}_{\pm n\rho}^{\prime}$ either
has the same components as $\vec{\mu}$ but shifted by $\pm n\rho$,
i.e. $\mu_{j}\to\mu_{j}\pm n\rho$, if $x$ coincides with a vertex
of the graph located at $x(v_{j})$, or it
will be $\vec{\mu}$ but with a new component $\pm n\rho$, i.e. will be $\{\ldots,\mu_{j},\pm n\rho,\mu_{j+1},\ldots\}$,
if $x_{v_{j}}<x<x_{v_{j+1}}$.

The final term to be considered is $K_{\varphi}^{2}E^{\varphi}$. For
it, we choose the representation proposed in \citep{Martin-Benito2008,Martin-Benito2009}, that is we define this operator as
\begin{align}\nonumber
&\hat{\Theta}(x)|g,\vec{k},\vec{\mu},M\rangle=\sum_{v_{j}\in g}\delta(x-x(v_{j}))\\
&\times\hat{\Omega}_{\varphi}^{2}(v_{j})|g,\vec{k},\vec{\mu},M\rangle,
\end{align}
where the non-diagonal operator $\hat{\Omega}_{\varphi}(v_{j})$ is
written as 
\begin{align}\nonumber
&\hat{\Omega}_{\varphi}(v_{j})=\frac{1}{4i\rho}|\widehat{E^{\varphi}}|^{1/4}\bigg[\widehat{\sgn(E^{\varphi})}\big(\widehat{N_{2\rho}^{\varphi}}-\widehat{N_{-2\rho}^{\varphi}}\big)\\
&+\big(\widehat{N_{2\rho}^{\varphi}}-\widehat{N_{-2\rho}^{\varphi}}\big)\widehat{\sgn(E^{\varphi})}\bigg]|\widehat{E^{\varphi}}|^{1/4}\Big|_{v_{j}}.
\end{align}
This shows that we need to also represent $\left|E^{\varphi}\right|^{1/4}$
and $\sgn(E^{\varphi})$. This can be achieved by means of the spectral
decomposition of $\widehat{E^{\varphi}}$ on $\mathscr{H}_{{\rm kin}}$
as 
\begin{align}
        \left|\widehat{E^{\varphi}}\right|^{1/4}(v_{j})|g,\vec{k},\vec{\mu},M\rangle= & |\mu_{j}|^{1/4}|g,\vec{k},\vec{\mu},M\rangle,\\
        \widehat{\sgn\big(E^{\varphi}(v_{j})\big)}|g,\vec{k},\vec{\mu},M\rangle & =\sgn(\mu_{j})|g,\vec{k},\vec{\mu},M\rangle.
\end{align}
Combining these yields a complete representation of our Hamiltonian constraint
on $\mathscr{H}_{{\rm kin}}$ as 
\begin{widetext}
\begin{equation}
        \hat{{\cal H}}(N)=\int dxN(x)\widehat{E^{x}}\left\{ \hat{\Theta}+\left(4\lambda^{2}\widehat{E^{\varphi}}\widehat{E^{x}}^{2}-2\lambda  G_2\hat{M}\widehat{E^{\varphi}}\widehat{E^{x}}\right)-\widehat{\left[\frac{[(E^{x})^{\prime}]^{2}}{E^{\varphi}}\right]}\right\} .\label{eq:Hconst-repres}
\end{equation}
\end{widetext}

\subsection{Hamiltonian constraint: singularity resolution and solutions\label{sub:Sing-resolv}}

\subsubsection{Relation between volume and singularity}

Our singularity resolution argument is based on having a zero volume
at some point (or region) classically or having a zero volume eigenvalue
for the quantum volume operator in quantum theory. In other words,
a vanishing volume (spectrum) at a point or region means we have a
singularity there. Here we give an argument supporting this statement
for a generic 2D metric (with generic lapse and shift).

A generic ADM decomposed 2D metric can be written as 
\begin{equation}
        g_{\mu\nu}=\left(\begin{array}{cc}
                -N^{2}+\left(N^{1}\right)^{2}q_{11} & -N^{1}q_{11}\\
                -N^{1}q_{11} & q_{11}
        \end{array}\right)\label{eq:gener-met}
\end{equation}
where $q_{11}$ is the spatial metric and $N$ and $N^{1}$ are lapse
and shift respectively. Since we have a one dimensional spatial hypersurface,
then 
\begin{equation}
        q_{11}=\det(q).\label{eq:q11-detq}
\end{equation}
Classically we have for the volume of a region $\textrm{R}$ in a spatial
hypersurface $\Sigma$, 
\begin{equation}
        V(\textrm{R})=\int_{\textrm{R}}dx\sqrt{\det(q)}.
\end{equation}
So if at some region we have $\det(q)=0$, this means that we will
get $V(\textrm{R})=0$ in that region. On the other hand, if $\det(q)=0$,
then due to (\ref{eq:gener-met}) and (\ref{eq:q11-detq}), we will
have for that region, a metric 
\begin{equation}
        g_{\mu\nu}=\left(\begin{array}{cc}
                -N^{2} & 0\\
                0 & 0
        \end{array}\right)
\end{equation}
independent of the lapse and shift. It turns out that  the Riemann
invariants of the above metric (in that region) blow up and thus we
have a singularity there. So, we conclude that in 2D, a vanishing volume
in a region means existence of singularity in that region. However,
this does not happen for a generic genuine  4D metric.

Now, for the quantum volume operator of the CGHS we have 
\begin{equation}
        \hat{\mathcal{V}}|\mathnormal{\mathnormal{g}},\vec{k},\vec{\mu},M\rangle\propto\sum_{v_{j}\in\mathnormal{g}}|\mu_{j}||\mathnormal{g},\vec{k},\vec{\mu},M\rangle,
\end{equation}
which means that a vanishing volume in a region corresponds to having
all the $\mu_{j}$'s equal to zero for that region (and not for the
whole spatial hypersurface). If we assume that the statement ``$V(\textrm{R})=0\Rightarrow$
singularity'', can be carried on to the quantum level, then we can
say that a region (or hypersurface) described by a state with none
of its $\mu_{j}$'s being zero is a region that does not contain any
singularity. This argument which to our knowledge only works generically
for genuine 2D spacetime metrics is the one we shall use to argue for
singularity resolution in the next subsection.

\subsubsection{Properties of the Hamiltonian constraint and singularity resolution\label{sub:prop}}

Keeping the argument of the previous subsection in mind and having
obtained a representation of the Hamiltonian constraint (\ref{eq:Hconst-repres})
on $\mathscr{H}_{{\rm kin}}$, we shall study some interesting properties
of this quantum Hamiltonian constraint. These properties will facilitate the identification
of the space of solutions of this constraint, and its relation with the singularity resolution it provides.

Let us consider any basis state $|g,\vec{k},\vec{\mu},M\rangle\in\mathscr{H}_{{\rm kin}}$.
It turns out that the action of this constraint on it yields 
\begin{widetext}
\begin{align}\nonumber
& \hat{{\cal H}}(N)|g,\vec{k},\vec{\mu},M\rangle=\sum_{v_{j}\in g}\bigg(N(x_{j})(\hbar G_2k_{j}) \\
& \times\Big[f_{0}(\mu_{j},k_{j},M)|g,\vec{k},\vec{\mu},M\rangle-f_{+}(\mu_{j})|g,\vec{k},\vec{\mu}_{+4\rho_{j}},M\rangle-f_{-}(\mu_{j})|g,\vec{k},\vec{\mu}_{-4\rho_{j}},M\rangle\Big]\bigg),\label{eq:H-const-act}
\end{align}
where the functions $f$ read 
\begin{align}
& f_{\pm}(\mu_{j})= \frac{\hbar G_{2}}{16\rho^{2}}|\mu_{j}|^{1/4}|\mu_{j}\pm2\rho|^{1/2}|\mu_{j}\pm4\rho|^{1/4} \left[\sgn(\mu_{j}\pm4\rho)+\sgn(\mu_{j}\pm2\rho)\right]\left[\sgn(\mu_{j}\pm2\rho)+\sgn(\mu_{j})\right],\label{eq:fpm}\\
&f_{0}(\mu_{j},k_{j},k_{j-1},M)= (\hbar G_{2})^{3}\lambda^{2}\left(1-\frac{G_{2}\hat{M}}{2\hbar G_{2}k_{j}\lambda}\right)\mu_{j}k_{j}^{2}-\frac{\hbar G_{2}}{\rho^{2}}\left(|\mu_{j}+\rho|^{1/2}-|\mu_{j}-\rho|^{1/2}\right)^{2}[k_{j}-k_{j-1}]^{2}\nonumber\\
&+\frac{\hbar G_{2}}{16\rho^{2}}\left\{ |\mu_{j}|^{1/2}|\mu_{j}+2\rho|^{1/2}\left[\sgn(\mu_{j})+\sgn(\mu_{j}+2\rho)\right]^{2}+|\mu_{j}|^{1/2}|\mu_{j}-2\rho|^{1/2}\left[\sgn(\mu_{j})+\sgn(\mu_{j}-2\rho)\right]^{2}\right\}\label{eq:f0}
\end{align}
\end{widetext}
Looking at (\ref{eq:H-const-act}) and the form of (\ref{eq:fpm})
and (\ref{eq:f0}), we notice some important points: 
\begin{enumerate}
        \item The scalar constraint admits a natural decomposition on each vertex
        $v_{j}$, such that it can be regarded as a sequence of quantum operators
        acting almost independently on them, up to the factors $\Delta k_{j}=k_{j}-k_{j-1}$.
        In other words, there would not be coupling among different vertices
        if it were not for the factor $\Delta k_{j}$. 
        \item The number of vertices on a given graph $g$ is preserved under the
        action of the Hamiltonian constraint. 
        \item \label{item:point-k0} The constraint (\ref{eq:H-const-act}) leaves
         the sequence of integers $\{k_{j}\}$ of each graph $g$ invariant.
        For instance, if we consider a ket $|g,\vec{k},\vec{\mu},M\rangle$,
        the successive action of the scalar constraint on it generates a subspace
        characterized by the original quantum numbers $\vec{k}$. 
        \item \label{item:point-mu0} The restriction of the constraint to any vertex
        $v_{j}$ acts as a difference operator mixing the real numbers $\mu_{j}$. In this case, this difference operator
         only relates those states which have $\mu_{j}$'s that belong to 
        a semi-lattices  of step $4\rho$ due to the  form of $f_\pm(\mu_j)$, 
        that vanishes in the intervals $[0,\mp 2\rho]$.
        \item \label{point-5} Starting from a state for which none of  $\mu_{j}$'s are zero (i.e.
        a state containing no singularity), the result of the action of the constraint never leaves us in
        a state with any of $\mu_{j}$'s being zero (also look at the details
        in section \ref{sub:properties-sol}). 
\end{enumerate}

The point number \ref{point-5}, which maybe is the most important of these, states that the subspace of $\mathscr{H}_{{\rm kin}}$ containing spin networks for which no $\mu_{j}$ is zero is preserved under the action of the Hamiltonian constraint. Simply put, if one originally starts
with a state with no singularity  (in the sense of $\mu_j=0$), then one will
never end up in a state containing a singularity. Analogous arguments could be applied to the $k_j$ quantum numbers, however as mentioned above, $k_j$ are already preserved by the constraint (unlike the $\mu_j$ valences of the vertices).

Thus, one can restrict the study only to the subspace of $\mathscr{H}_{{\rm kin}}$
for which there is no $\mu_{j}=0$ and $k_{j}=0$.  As a result this restriction, we expect that also      in the physical Hilbert space,
we will never have any state with a singularity.

\subsection{Solutions to the Hamiltonian constraint and the physical Hilbert space\label{sub:properties-sol}}

Let us consider a generic solution, $\langle\Psi_{g}|$, to the Hamiltonian
constraint, i.e., a generic state annihilated by this constraint. Assuming
$\langle\Psi_{g}|$ belongs to the algebraic dual of the dense subspace
$Cyl$ on the kinematical Hilbert space, and that it can be written as 
\begin{align}\nonumber
&\langle\Psi_{g}|=\int_{0}^{\infty}dM\sum_{\vec{k}}\sum_{\vec{\mu}}\langle g,\vec{k},\vec{\mu},M|\psi(M)\chi(\vec{k})\\
&\times\phi(\vec{k},\vec{\mu},M),\label{eq:generic-kin-state}
\end{align}
then the annihilation by the Hamiltonian constraint dictates that 
\begin{equation}
\langle\Psi_{g}|\hat{{\cal H}}(N)^{\dagger}=\sum_{v_{j}\in g}\langle\Psi_{g}|N_{j}\mathfrak{\hat{H}}_{j}^{\dagger}=0,\label{eq:kin-state-dual-H}
\end{equation}
where $\hat{\mathfrak{H}}_{j}$ are difference operators acting on each vertex $v_j$ and $N_{j}=N(x_{j})$ is the lapse function evaluated on the corresponding vertex. In this case the functions $\phi(\vec{k},\vec{\mu},M)$ admits a natural
decomposition of the form 
\begin{equation}
        \phi(\vec{k},\vec{\mu},M)=\prod_{j=1}^{V}\phi_{j}(k_{j},k_{j-1},\mu_{j},M).
\end{equation}
One can then easily see that the solutions must fulfill, at each $v_{j}$, a difference equation of the form 
\begin{align}\nonumber
&-f_{+}(\mu_{j}-4\rho)\phi_{j}(k_{j},k_{j-1},\mu_{j}-4\rho,M)\\\nonumber
&-f_{-}(\mu_{j}+4\rho)\phi_{j}(k_{j},k_{j-1},\mu_{j}+4\rho,M)\\ 
&+f_{0}(k_{j},k_{j-1},\mu_{j},M)\phi_{j}(k_{j},k_{j-1},\mu_{j},M)=0,\label{eq:one-per-vj-diffc}
\end{align}
which is a set of difference equations to be solved together. We will provide a partial resolution of the problem by means of analytical considerations. All the details can be found in appendix \ref{sec:app-A}. Let us consider a particular vertex $v_j$. In the following we will omit any reference to the label of the vertex. Due to the property \ref{item:point-mu0}, where $\mu$ belongs to the semi-lattices of the form $\mu=\epsilon\pm4\rho{\rm n}$
where ${\rm n}\in\mathbb{N}$ and $\epsilon\in(0,4\rho]$, different orientations of $\mu$ are decoupled. Without loss of generality, we will restrict the study to a particular subspace labeled by $\epsilon$, unless otherwise specified. This shows that the Hamiltonian constraint only relates states belonging to separable subspaces of the original kinematical Hilbert space. 

These properties of the solutions together with their asymptotic limit $\mu\to\infty$, assuming the solutions are smooth there, will allow us to understand several aspects of the geometrical operators (under some assumptions about their spectral decomposition). More concretely, the solutions for $\mu\to\infty$ satisfy, up to a global factor $[(\hbar G_2)^2k]$, the differential 
equation 
\begin{align}\nonumber
&-4\mu\partial_{\mu}^{2}\phi-4\partial_{\mu}\phi-\frac{4\Delta k^{2}-1}{4\mu}\phi\\
&+\left(1-\frac{G_2M}{2\hbar G_2\lambda k}\right)(\hbar G_2\lambda)^{2}k^{2}\mu\phi=0,\label{eq:asympt-diff-op}
\end{align}
in a very good approximation if they are smooth functions of $\mu$. The last term plays the role of the square of a frequency of an harmonic oscillator. But the sign of this term depends on the concrete quantum numbers. Therefore, this equation admits both oscillatory solutions and exponentially growing or decreasing ones. More concretely, this differential equation is a modified Bessel equation if the sign of its last coefficient is positive, i.e. $k<M/2\hbar\lambda$, and a Bessel equation whenever that coefficient is negative, i.e. $k>M/2\hbar\lambda$. In Appendix \ref{sec:app-A} we include the details about the properties of the solutions in these two different regimes. Let us summarize
the results obtained there: 
\begin{itemize}
        \item For $k<M/2\hbar\lambda$, the
                Hamiltonian constraint takes the form 
                \begin{equation}
                \omega+\left(1-\frac{G_2\hat{M}}{2\hbar G_2\lambda                 k}\right)(\hbar G_2\lambda)^{2}k^{2}=0.\label{eq:scalar-constr-1}
                \end{equation}
                where $\omega$ is the positive eigenvalue of the difference
                operator of (\ref{eq:int-diff-op}) that belongs to its continuous
                spectrum and which is non-degenerate. The corresponding eigenfunction
                $|\phi_{\omega}^{{\rm cnt}}\rangle$ behaves as an exact standing wave
                in $\mu$ of frequency $\sigma(\omega)$ in the limit $\mu\to\infty$.
        \item On the other hand, for $k>M/2\hbar\lambda$,
                the constraint is simply 
                \begin{equation}
                \omega_n(M,k,\epsilon)-\Delta k^{2}=0,\label{eq:scalar-constr-2}
                \end{equation}
                where, again, $\omega_n$ is the positive eigenvalue
                of the difference operator defined in (\ref{eq:diff-op-ext}), but
                this time it belongs to its discrete spectrum and is also non-degenerated.
                The corresponding eigenstates $|\phi_{n}^{{\rm dsc}}\rangle$, with
                $n\in\mathbb{N}$, emerges out of $\mu\simeq\epsilon$, grow exponentially
                until reach an stable regime, and at some $\mu\simeq\mu_{r}$ the
                eigenfunction enters in a classically forbidden region and decays
                exponentially (see \citep{cfrw} for a related treatment). Besides,
                the eigenfunctions $\omega_{n}$ form a discrete sequence
                of real numbers, all of them depending continuously on the parameter
                $\epsilon$. This dependence is crucial in order to have a consistent
                constraint solution, since the sequence of discrete $\Delta k^{2}$
                is not expected to coincide with the sequence of $\omega_{n}$
                for a global fixed $\epsilon$. Therefore, we expect that the parameter
                $\epsilon$ must be conveniently modified according to the values
                of $M$, $k$ and the constraint equation (\ref{eq:scalar-constr-2}). 
\end{itemize}

These previous results have not been confirmed numerically
(as well as those of \citep{R.Gambini2013}), though they will be
a matter of future research. Let us comment, however, that they are
based on very robust, previous results on different scenarios already
studied in the LQC literature (see \citep{Martin-Benito2009,cfrw,aps-old,aps-imp,acp}).
Therefore, unless a very subtle point comes into play, the
mentioned properties are expected to be fulfilled.

In a final step, one should build the physical Hilbert space.
The states belonging to this space are the ones that admit the symmetries
of the model, i.e. the states which are invariant under both the Hamiltonian
and diffeomorphism constraints. As usual in LQG, one applies the group
averaging technique to get these states and the induced inner product
on the resultant subspace that is provided by this process. One can
start with the Hamiltonian constraint. Then the states averaged by
members of the group associated to the Hamiltonian constraint are
\begin{align}\nonumber
&\langle\Psi_{g}^{\mathcal{H}}|=\int_{-\infty}^{\infty}\prod_{n=1}^{V}d\mathbf{g}_{n}\int_{0}^{\infty}dM\sum_{\vec{k}}\sum_{\vec{\mu}}\langle g,\vec{k},\vec{\mu},M|\\
&\times \psi(M)\chi(\vec{k})\phi(\vec{k},\vec{\mu},M),\label{eq:gr-av-gen}
\end{align}
where 
\begin{equation}
        \mathbf{g}=e^{i\mathfrak{g}}
\end{equation}
is the group member associated to the member of the Lie algebra $\mathfrak{g}$.
In the case of the algebra member, being the Hamiltonian constraint
$\hat{\mathcal{H}}(N_{j})=\sum_{v_{j}}N_{j}\hat{\mathfrak{H}}_{j}$,
we have 
\begin{equation}
        \mathfrak{g}=\hat{\mathcal{H}}(N_{l})=\sum_{v_{l}}N_{l}\hat{\mathfrak{H}}_{l}.
\end{equation}
Thus in this case, from (\ref{eq:gr-av-gen}) we get for the group
averaged state 
\begin{widetext}
\begin{align}
        \langle\Psi_{g}^{\mathcal{H}}|= & \frac{1}{(2\pi)^V}\int_{-\infty}^{\infty}de^{iN_{1}\hat{\mathfrak{H}}_{1}}\ldots\int_{-\infty}^{\infty}de^{iN_{V}\hat{\mathfrak{H}}_{V}}\int_{0}^{\infty}dM\sum_{\vec{k}}\sum_{\vec{\mu}}\langle g,\vec{k},\vec{\mu},M|\psi(\vec{k},\vec{\mu},M)\nonumber \\
        = & \frac{1}{(2\pi)^V}\int_{-\infty}^{\infty}dN_{1}\ldots\int_{-\infty}^{\infty}dN_{V}\exp\left[i\sum_{n=1}^{V}N_{n}\hat{\mathfrak{H}}_{n}\right]\int_{0}^{\infty}dM\sum_{\vec{k}}\sum_{\vec{\mu}}\langle g,\vec{k},\vec{\mu},M|\psi(\vec{k},\vec{\mu},M)\label{eq:gr-av-gen-explic}
\end{align}
\end{widetext}
The final states are endowed with a suitable inner product defined as
\begin{align}
\|\Psi_{g}^{\mathcal{H}}\|^2=\langle\Psi_{g}^{\mathcal{H}}|\Psi_{g}\rangle,
\end{align}
where the ket belongs to the kinematical Hilbert space and the bra is the corresponding state after being averaged with the Hamiltonian constraint. In order to obtain explicitly the inner product, we may write $|\Psi_{g}\rangle$ in the basis of states of the geometrical operators involving the scalar constraint (see App. \ref{sec:app-A}). In this case 
\begin{align}\nonumber
&\langle\Psi_{g}^{\mathcal{H}}|\Psi_{g}\rangle =\int_{0}^{\infty}dM\sum_{\vec{k}}\int d\omega_1\ldots d\omega_V \\
&\times\prod_{j =1}^{V}\delta\Big(\omega_j-F(k_j,M)\Big)|\psi(\vec{k},\vec{\omega},M)|^2,
\end{align}
where $F(k_j,M)$, at each vertex $v_j$, is given by the last addend in the left hand side of Eqs. (\ref{eq:disc-const}) or (\ref{eq:cont-const}) depending if $\left(k_j-M/2\hbar\lambda\right)$ is positive or negative, respectively, i.e.
\begin{align}
F(k_j,M)=&(\Delta k_j)^2\quad {\rm if}\quad k_j>M/2\hbar\lambda,\\ \nonumber
F(k_j,M)=&\left(1-\frac{G_2M}{2\hbar G_2\lambda k_j}\right)(\hbar G_2\lambda)^{2}k_j^{2}
\end{align}
otherwise. The final step is to construct the solutions to the Hamiltonian constraint which are invariant under the spatial diffeomorphisms (generated by the diffeomorphism constraint). In this case we follow the ideas of the full theory \cite{diff-inv}. There, one constructs a rigging map from the original Hilbert space to the space of diffeomorphism invariant states by averaging the initial states with respect to the group of finite diffeomorphisms. The resulting averaged states are a superposition of the original states but with their vertices in all possible positions in the original 1-dimensional manifold, but preserving the order of the edges and vertices. So, a physical state will be
\begin{equation}
\langle\Psi^{\rm phys}|=\sum_{ g\in[g]} \langle\Psi_{g}^{\mathcal{H}}|
\end{equation}
and the inner product is then
\begin{equation}\label{eq:phys:inner-p}
\|\Psi^{\rm phys}\|^2=\langle\Psi^{\rm phys}|\Psi_g\rangle,
\end{equation}
where, again, the ket belongs to the kinematical Hilbert space and the bra is the physical solution. In the last product, only a finite number of finite terms contribute, for all $|\Psi_g\rangle$ in the kinematical Hilbert space, so the inner product is finite and well defined. Let us mention that the diffeomorphism invariance of the inner product is guaranteed since if we compute Eq. (\ref{eq:phys:inner-p}) with any other state related to $|\Psi_g\rangle$ by a spatial diffeomorphism, it would yield exactly the same result. For a recent discussion see \cite{bojo-spher}.

At the end of this process, we are left with a vector space of states that are invariant under both constraints, and an inner product on
this space, induced by the group averaging processes, rendering
this vector space a Hilbert space. The resultant Hilbert space of
diffeomorphism invariant states are the equivalence classes of diffeomorphism
invariant graphs $[\mathnormal{g}]$ solutions to the scalar constraint.

Let us conclude with some remarks. In the classical
theory, the geometry possesses a singularity whenever the determinant
of the metric $q$ 
vanishes at some point.
In this manuscript, the vanishing of $q$ 
corresponds to the vanishing of $E^{\varphi}$ 
at a given region (local singularity). In the quantum theory we can
find an analogous situation, for instance if a graph $g$ has $\mu_{j}=0$
at one or some given vertices. Fortunately, the
quantum theory allows us to avoid these undesired divergences. The
key idea consists in identifying a suitable invariant domain of the
scalar constraint, free of such states with nonvanishing $\mu_{j}$.
In this way, the solutions to the constraints will have
support only on them, preventing the vanishing of $\mu_{j}$ (and $k_{j}$)
at any vertex. It is straightforward to prove, as a direct consequence
of the previous points \ref{item:point-k0} and \ref{item:point-mu0},
that the subspace formed by kets such that their sequences $\{\mu_{j}\}$
(and $\{k_{j}\}$) contains no vanishing components, remains invariant
under the action of the Hamiltonian constraint (\ref{eq:H-const-act}). In particular,
point \ref{item:point-mu0} tells us that we can never reach a vanishing
$\mu_{j}$ by successive action of the scalar constraint, and point
\ref{item:point-k0} tells us that any sequence of $\{k_{j}\}$ will
remain invariant. In conclusion, the restriction to this invariant
domain allows us to resolve the classical singularity.

However, given that the sequences $\{k_{j}\}$ are unaltered
by the scalar constraint, and since they apparently have no significance in singularity resolution of this model, 
we do not see any fundamental argument
for discriminating those with vanishing $\{k_{j}\}$ components with respect to
the remaining ones. In \citep{Gambini2013a} it was suggested that
the reality conditions of some observables of the model provides a
quantum theory free of singularities. However, due to some important differences of 
that model with the present one, we have not been able
to identify such suitable observables in our model along those lines.

\subsection{Quantum observables\label{sub:observables}}

We saw in section \ref{sub:properties-sol} that the Hamiltonian constraint
does not create any new vertices in the graph $\mathnormal{g}$ on which
it acts (and obviously neither does the diffeomorphism constraint).
This means that there is a Dirac observable $\hat{\mathcal{N}}_{v}$
in the bulk corresponding to the fixed number of vertices $\mathcal{N}_{v}=V$
of a graph $\mathnormal{g}$, 
\begin{equation}
        \hat{\mathcal{N}}_{v}\Psi_{\textrm{phys}}=\mathcal{N}_{v}\Psi_{\textrm{phys}}.
\end{equation}
This observable is strictly quantum and has no counterpart in the
classical theory.

On the other hand, since this model has only one spatial direction,
under the action of the diffeomorphism constraint the points
can not pass each other, i.e., the order of the positions of the vertices
is preserved. This means that, associated to this preservation, we
can identify another new strictly quantum observable in the bulk,
$\hat{O}(z)$ such that 
\begin{align}
        \hat{O}(z)\Psi_{{\rm phys}}=k_{\Int(z\mathcal{N}_{v})}\Psi_{{\rm phys}}, &  & z\in[0,1]\label{eq:obs-order}
\end{align}
where $\Int(z\mathcal{N}_{v})$ is the integer part of $z\mathcal{N}_{v}$. Together with them, we also have the observable corresponding to the mass $\hat M$, which does have an analogous classical Dirac observable.

Besides, as it was first observed by the authors of Ref.  \citep{Gambini2013a,R.Gambini2013},
one can construct an evolving constant associated to $E^{x}$ from
the above observable as 
\begin{align}
        \widehat{E^{x}}(x)\Psi_{{\rm phys}}=\hbar G_2\hat{O}(z(x))\Psi_{{\rm phys}},
\end{align}
with $z(x):[0,x]\rightarrow[0,1]$. Since $E^{x}$ has classical and quantum mechanically a different interpretation in the CGHS model
than in 3+1 spherical symmetry, i.e., in the former it is related to the
dilaton field, one should also take caution about its interpretation.

These two observables were first introduced for the 3+1 spherically
symmetric case in \citep{Gambini2013a} and due to the similarities
of the two models, we can see that they exist also for the CGHS model.
Particularly, the observable in (\ref{eq:obs-order}) arises due to
the existence of only one (radial) direction in both cases. So one
can expect that such a quantum observable will exist in many genuinely
2D and symmetry-reduced models with only one radial direction
in which the quantum theory implements the spatial diffeomorphism symmetry as in loop quantum gravity.

It is worth commenting that one can promote the metric component $\hat E^\varphi$ as a parameterized observable. For it, we can choose the phase space variable $K_\varphi$ as an internal time function (of parametric function). Moreover, by means of the Hamiltonian constraint (on shell), it is possible to define the parameterized observable
\begin{widetext}
\begin{equation}
\hat E^{\varphi}(x)\Psi_{{\rm phys}}=\frac{\partial_x\widehat{E^{x}}(x)}{\sqrt{4[\widehat{E^{x}}(x)]^{2}\lambda^{2}+\frac{\sin^{2}(\rho K_{\varphi})}{\rho^2}-2\lambda G_2\hat M\widehat{E^{x}}(x)}}\Psi_{{\rm phys}},
\end{equation}
\end{widetext}
which is defined in terms of the parameter functions $z(x)$ and $K_\varphi(x)$, and the observables $\hat M$ and $\hat{O}$ (through the definition of $\widehat{E^{x}}(x)$).

\section{Summary and conclusion}

We have shown that, with the introduction of polar-type
variables for a CGHS dilatonic black hole and a rewrite of its Hamiltonian 
in terms of those
variables, one can follow recent LQG inspired methods, first introduced in 3+1 spherically-symmetric case --also written in 
polar-type variables--, to remove the singularity of the CGHS
model. The proposal is based on the assumption (proven here for the case
of a 2D generic metric) that states with zero volume are those
containing a spacetime singularity. Then, singularity resolution follows if one can show
that if one starts from a state without a zero volume present in it, one can restrict the evolution
to a subspace of the Hilbert space that contains no zero volume states.
In other words, the subspace of quantum spacetime states without  a singularity is preserved under the action of the quantum Hamiltonian constraint. 

This analysis may be extended further when a matter field is present in
the theory and one might then study the backreaction, but that analysis will 
certainly be more involved and is outside the scope of this paper.
Although it has been shown recently \cite{CGHS-CL-New} that even in the presence of  
matter (more precisely, massless scalar field), one can get a Lie algebra of constraints 
by strong Abelianization of the $\{\mathcal{H}(N),\mathcal{H}(M)\}$ part of the classical 
constraint algebra, it is not clear whether the quantum theory is anomaly-free and also if 
one can get some useful information about the Hamiltonian constraint, as was possible in 
the present case without matter. Furthermore, the representation
of this constraint on the Hilbert space is expected to be much more involved. For a 
minimally-coupled scalar field (in the classical theory its dynamics reduces to the one of 
a scalar field in Minkowski) one can expect a more treatable model with respect to its 
analogue in 3+1 spherically-symmetric spacetimes, regarding its solubility. Nevertheless, 
this is an interesting future project worth pursuing, as is the study the Hawking 
radiation based on these results.

In any case, the analysis here presented must be viewed as a first step that requires further  
understanding, analysis and level of precision. It can hopefully be further extended
to give more insights on generic black hole singularity resolution
and, more generally, on quantum gravity itself. 

\begin{acknowledgments}
We would like to thank J.D. Reyes for discussions and comments. 
A.C. was in part supported by DGAPA-UNAM IN103610 
grant, by CONACyT 0177840 and 0232902 grants, by the PASPA-DGAPA program, by NSF PHY-1505411 and 
PHY-1403943 grants, and by the Eberly Research Funds of Penn State.
S.R. would like to acknowledge the support from the Programa de Becas Posdoctorales, Centro de 
Ciencias Matematicas, Campos Morelia, UNAM and DGAPA, partial support of CONACyT grant number 
237351: Implicaciones F\'{i}sicas de la Estructura del Espaciotiempo, the support of the the 
PROMEP postdoctoral fellowship (through UAM-I) and the grant from Sistema Nacional de 
Investigadores of CONACyT. J.O. acknowledges funds by Pedeciba, grant FIS2014-54800-C2-2-P 
(Spain) and grant NSF-PHY-1305000 (USA). 

\end{acknowledgments}

\appendix

\section{Spectrum of geometrical operators \label{sec:app-A}}

In this appendix we will discuss some properties
of the Hamiltonian constraint restricted to one arbitrary vertex $v_j$ (we will omit the label $j$ in the following). Let us recall
that the local scalar constraint of this model, once promoted to a
quantum operator, acts (almost) independently on each vertex. Its
action on the corresponding states is 
\begin{align}\nonumber
\hat{\mathfrak{H}}|g,k,\mu,M\rangle=&(\hbar G_2k) \Big[f_{0}(\mu,k,M)|g,k,\mu,M\rangle\\ \nonumber
&-f_{+}(\mu)|g,k,\mu+4\rho,M\rangle\\
&-f_{-}(\mu)|g,k,\mu-4\rho,M\rangle\Big],\label{eq:H-1v-const-act}
\end{align}
\begin{widetext}
with the functions 
\begin{align}
&f_{\pm}(\mu)=  \frac{\hbar G_2}{16\rho^{2}}|\mu|^{1/4}|\mu\pm2\rho|^{1/2}|\mu\pm4\rho|^{1/4}\left[\sgn(\mu\pm4\rho)+\sgn(\mu\pm2\rho)\right]\left[\sgn(\mu\pm2\rho)+\sgn(\mu)\right],\label{eq:fpm-1vx}\\
&f_{0}(\mu,k,M)=  \lambda^{2}\left(1-\frac{G_2M}{2\hbar G_2k\lambda}\right)(\hbar G_2)^{3}\mu k^2+\frac{\hbar G_2}{16\rho^{2}}\left\{ |\mu|^{1/2}|\mu+2\rho|^{1/2}\left[\sgn(\mu)+\sgn(\mu+2\rho)\right]^{2}\right.\nonumber \\
& \left.+|\mu|^{1/2}|\mu-2\rho|^{1/2}\left[\sgn(\mu)+\sgn(\mu-2\rho)\right]^{2}\right\} -\frac{\hbar G_2}{\rho^{2}}\left(|\mu+\rho|^{1/2}-|\mu-\rho|^{1/2}\right)^{2}\Delta k^{2}.\label{eq:f0-1vx}
\end{align}
\end{widetext}
Here, $\Delta k$ is proportional to the
eigenvalue of the operator $\widehat{(E^{x}(x))'}$. This operator
will be diagonal on the spin network basis of states as well as its
explicit form will depend on the definition of the operator $\hat E^x$.

The action of the scalar constraint resembles the
one of a second order difference operator since it relates three consecutive
points in a lattice with constant step. The consequence is that any
function $\phi(k,\mu,M)$ that is solution to the equation $(\phi|\hat{\mathfrak{H}}^{\dagger}=0$
has support on lattices of step $4\rho$, as we can deduce by direct
inspection of Eq.~\eqref{eq:H-1v-const-act}. Moreover, due to the
functions $\left[\sgn(\mu\pm2\rho)+\sgn(\mu)\right]$ in (\ref{eq:fpm-1vx}),
$f_{\pm}(\mu)$ vanishes on $[0,\mp2\rho]$ respectively. Thus different
orientations of $\mu$ are decoupled by the difference operator \eqref{eq:H-1v-const-act}.
We conclude that $\mu$ belongs to semi-lattices of the form $\mu=\epsilon\pm4\rho{\rm n}$
where ${\rm n}\in\mathbb{N}$ and $\epsilon\in(0,4\rho]$. Without
loss of generality, we will restrict the study to a particular subspace
labeled by $\epsilon$, unless otherwise specified. This shows that
the Hamiltonian constraint only relates states belonging to separable
subspaces of the original kinematical Hilbert space.

The solutions $\phi(k,\mu,M)$
fulfill the equation
\begin{align}\nonumber
&-f_{+}(\mu-4\rho)\phi(k,\mu-4\rho,M)\\ \nonumber
&-f_{-}(\mu+4\rho)\phi(k,\mu+4\rho,M)\\
&+f_{0}(k,\mu,M)\phi(k,\mu,M)=0.\label{eq:1vs-diffc}
\end{align}
One can straightforwardly realize that, for any choice
of the initial triad section $\mu=\epsilon$, they are completely determined by their initial data $\phi(k,\mu=\epsilon,M)$.
In particular, our difference operator evaluated at $\mu=\epsilon$
relates the solution coefficient $\phi(k,\mu=\epsilon+4\rho,M)$ with
only the initial data $\phi(k,\mu=\epsilon,M)$, which can be solved
easily. Therefore, the difference equation evaluated at the next successive
lattice points can also be solved straightforwardly, once the initial
data $\phi(k,\mu=\epsilon,M)$ is provided. Without loss of generality,
we will fix it to be real. This allows us to conclude that, since
the coefficients of the corresponding difference equation (\ref{eq:1vs-diffc})
are also real functions, the solutions $\phi(k,\mu,M)$ at any triad
section $\mu=\epsilon\pm4\rho{\rm n}$ will be also real functions.

Besides, the solutions to Eq. (\ref{eq:1vs-diffc}),
for constant values of the quantum numbers $k$, $M$, and the cosmological
constant $\lambda$, have different asymptotic limits
at $\mu\to\infty$. Concretely, if they fulfill $k<M/2\hbar\lambda$
or $k>M/2\hbar\lambda$, the physically relevant
solutions either oscillate or decay exponentially,
respectively, in that limit. 

We will focus now in the study of the solutions in
the cases in which $k>M/2\hbar\lambda$. In this
regime, it is more convenient to carry out a transformation in the
functional space of solutions in order to achieve a suitable separable
form of the constraint equation. In particular, following the ideas
of Ref.~\cite{Martin-Benito2009}, we will introduce a bijection
on the space of solutions defined by the scaling of the solutions
\begin{equation}
\phi^{{\rm dcr}}(k,\mu,M)=(\hbar G_2)^{1/2}\hat{b}(\mu)\phi(k,\mu,M),\label{eq:scaling-out}
\end{equation}
with 
\begin{equation}
\hat{b}(\mu)=\frac{1}{\rho}(|\hat{\mu}+\rho|^{1/2}-|\hat{\mu}-\rho|^{1/2}).\label{eq:bmu-coeff}
\end{equation}
We might notice that the functions $\hat{b}(\mu)$ only vanish for
$\mu=0$. But this sector has been decoupled, since $\mu$ belong
to semi-lattices with a global minimum at $\mu=\epsilon>0$. Therefore,
the function $\hat{b}(\mu)$ never vanishes and the previous scaling
is invertible. The new functions $\phi^{{\rm dcr}}(k,\mu,M)$ now
fulfill the difference equation 
\begin{align}\nonumber
-f_{+}^{{\rm dcr}}(\mu-4\rho)\phi^{{\rm dcr}}(k,\mu-4\rho,M)\\ \nonumber
-f_{-}^{{\rm dcr}}(\mu+4\rho)\phi^{{\rm dcr}}(k,\mu+4\rho,M)\\
+f_{0}^{{\rm dcr}}(k,\mu,M)\phi^{{\rm dcr}}(k,\mu,M)=0.\label{eq:1vs-dcr-diffc}
\end{align}
where the new coefficients are now 
\begin{widetext}
\begin{align}
f_{\pm}^{{\rm dcr}}(\mu)= & \frac{1}{16\rho^{2}b(\mu)b(\mu\pm4\rho)}|\mu|^{1/4}|\mu\pm2\rho|^{1/2}|\mu\pm4\rho|^{1/4}\left[\sgn(\mu\pm4\rho)+\sgn(\mu\pm2\rho)\right]\nonumber\\
&\times\left[\sgn(\mu\pm2\rho)+\sgn(\mu)\right],\label{eq:f+-densitized}\\
f_{0}^{{\rm dcr}}(\mu,k,M)= & \frac{\mu}{b(\mu)^{2}}\left(1-\frac{G_2M}{2\hbar G_2\lambda k}\right)(\hbar G_2\lambda)^{2}k^{2}+\frac{1}{16\rho^{2}b(\mu)^{2}}\left[(|\mu||\mu+2\rho|)^{1/2}\left[\sgn(\mu)+\sgn(\mu+2\rho)\right]^{2}\right.\nonumber \\
& \left.+(|\mu||\mu-2\rho|)^{1/2}\left[\sgn(\mu)+\sgn(\mu-2\rho)\right]^{2}\right]-\Delta k^{2},
\end{align}
\end{widetext}
This difference operator can be naively interpreted as a densitized scalar
constraint, for instance, like the one emerging after choosing the
lapse function $N\hat{b}(\mu)^{-2}$ (together a suitable factor ordering and a global factor $\hbar G_2$).
Let us denote this scalar constraint in the original scaling
by $\hat{\mathfrak{H}}$, and the corresponding scalar constraint
by $\hat{\mathfrak{H}}^{{\rm dcr}}$ in the new one. Both are related
by 
\begin{equation}
\hat{\mathfrak{H}}^{{\rm dcr}}=\hat{b}(\mu)^{-1}\hat{\mathfrak{H}}\hat{b}(\mu)^{-1}.\label{eq:hdcr-to-h}
\end{equation}
Now, we will study the difference operator 
\begin{equation}
\hat{\mathfrak{h}}^{{\rm dcr}}=\hat{\mathfrak{H}}^{{\rm dcr}}+\Delta k^{2}.
\end{equation}
We can deduce several properties about the spectrum of this difference
operator as well as of its eigenfunctions. Let us consider, for consistency,
its positive spectrum. The eigenvalue problem 
\begin{equation}
\hat{\mathfrak{h}}^{{\rm dcr}}|\phi_{\omega}^{{\rm dcr}}\rangle=\omega|\phi_{\omega}^{{\rm dcr}}\rangle.\label{eq:diff-op-ext}
\end{equation}
corresponds to a difference equation similar to equation \eqref{eq:1vs-dcr-diffc}
but with functions
\begin{widetext}
\begin{align}
\tilde{f}_{\pm}^{{\rm dcr}}(\mu)= & \frac{1}{16\rho^{2}b(\mu)b(\mu\pm4\rho)}|\mu|^{1/4}|\mu\pm2\rho|^{1/2}|\mu\pm4\rho|^{1/4}\left[\sgn(\mu\pm4\rho)+\sgn(\mu\pm2\rho)\right]\nonumber\\
&\times\left[\sgn(\mu\pm2\rho)+\sgn(\mu)\right],\\
\tilde{f}_{0}^{{\rm dcr}}(\mu,k,M,\omega)= & \frac{\mu}{b(\mu)^{2}}\left(1-\frac{G_2M}{2\hbar G_2\lambda k}\right)(\hbar G_2\lambda)^{2}k^{2}+\frac{1}{16\rho^{2}b(\mu)^{2}}\left[(|\mu||\mu+2\rho|)^{1/2}\left[\sgn(\mu)+\sgn(\mu+2\rho)\right]^{2}\right.\nonumber \\
& \left.+(|\mu||\mu-2\rho|)^{1/2}\left[\sgn(\mu)+\sgn(\mu-2\rho)\right]^{2}\right]-\omega.
\end{align}
\end{widetext}

Let us assume that the solutions to this difference
equation has a well defined and smooth limit $\mu\to\infty$. For practical purposes
this limit is similar to the limit $\rho\to0$, but keeping in mind
that while the former is expected to be well defined in our quantum
theory, the latter is not. This assumption involves that the solutions
$\phi_{\omega}^{{\rm dcr}}(\mu)$ must be continuous functions of
$\mu$. But this is not true for scales $\Delta\mu$ of the order of $4\rho$ (in the
previous asymptotic limit). This must be tested carefully,
but we will not deal with this question by now. We assume its validity, at least for eigenvalues with typical scales much bigger than $4\rho$.

Within this asymptotic regime and approximation, the solutions to the
previous difference equation \eqref{eq:diff-op-ext} satisfy in a
very good approximation the differential equation 
\begin{widetext}
\begin{align}
 0=&-\tilde{f}_{+}^{{\rm dcr}}(\mu-4\rho)\phi_{\omega}^{{\rm dcr}}(k,\mu-4\rho,M)-\tilde{f}_{-}^{{\rm dcr}}(\mu+4\rho)\phi_{\omega}^{{\rm dcr}}(k,\mu+4\rho,M)+\tilde{f}_{0}^{{\rm dcr}}(k,\mu,M,\omega)\phi_{\omega}^{{\rm dcr}}(k,\mu,M)\nonumber \\
=& -4\mu^{2}\partial_{\mu}^{2}\phi_{\omega}^{{\rm dcr}}(k,\mu,M)-8\mu\partial_{\mu}\phi_{\omega}^{{\rm dcr}}(k,\mu,M)+\left[\left(1-\frac{G_2M}{2\hbar G_2\lambda k}\right)(\hbar G_2\lambda)^{2}k^{2}\mu^{2}-\gamma^{2}\right]\phi_{\omega}^{{\rm dcr}}(k,\mu,M)\nonumber\\
&+{\cal O}(\rho^{2}/\mu^{2}),\label{eq:mod-bessel-diff-op}
\end{align}
\end{widetext}
with $\gamma^{2}=\omega+3/4$. Let us recall that this differential
equation can be analogously achieved if instead of adopting a loop
quantization, one adheres to a WDW representation for this setting,
with a suitable factor ordering. It corresponds to modified Bessel
equation, where its solutions are combinations of modified Bessel
functions of the form 
\begin{align}\nonumber
\lim_{\mu\to\infty}\phi_{\omega}^{{\rm dcr}}(k,\mu,M)=&Ax^{-1/2}{\cal K}_{i\gamma}\left(x\right)\\
&+Bx^{-1/2}{\cal I}_{i\gamma}\left(x\right),
\end{align}
with 
\begin{equation}
x=\mu\frac{\hbar G_2\lambda k}{2}\sqrt{\left(1-\frac{G_2M}{2\hbar G_2\lambda k}\right)}.
\end{equation}
In the limit $\mu\to\infty$, the solutions ${\cal I}$ and ${\cal K}$
grow and decay exponentially, respectively. Therefore, the latter
is the only contribution to the spectral decomposition of $\hat{\mathfrak{h}}^{{\rm dcr}}$.
In consequence, its possible (positive) eigenvalues $\omega$ are
non-degenerate. Besides, the functions ${\cal K}_{i\gamma}(x)$ are
normalized to 
\begin{equation}
\langle{\cal K}_{i\gamma}|{\cal K}_{i\gamma'}\rangle=\delta(\gamma-\gamma'),\label{eq:mbessel-norm}
\end{equation}
in $L^{2}(\mathbb{R},x^{-1}dx)$, since the normalization in this
case is ruled by the behavior of ${\cal K}_{i\gamma}(x)$ in the limit
$x\to0$, which corresponds to 
\begin{equation}
\lim_{x\to0}{\cal K}_{i\gamma}(x)\to A\cos\left(\gamma\ln|x|\right).\label{eq:mbessel-x0-lim}
\end{equation}
For additional details see, for instance, Ref. \citep{kiefer}. This
result is fulfilled in the continuous theory, whenever (\ref{eq:mod-bessel-diff-op})
is valid globally. But let us recall that we are dealing with a difference
equation possessing, in a good approximation, a continuous $\mu\to\infty$
limit, but not at all for $\mu\to0$. Therefore, the previous normalization
\eqref{eq:mbessel-norm} and the asymptotic limit \eqref{eq:mbessel-x0-lim}
have no meaning in our discrete theory. In this case, and in the absence
of a meticulous numerical study of the solutions of this equation,
we can only infer some properties about $\phi_{\omega}^{{\rm dcr}}(k,\mu,M)$.
One can convince oneself that our difference equation is similar to
the one studied in \citep{cfrw} for a closed FRW spacetime. In particular,
the eigenfunctions of such a difference operator have a similar asymptotic
behavior for $v\to\infty$ (or equivalently $\mu\to\infty$ in our
model). Nevertheless, the spectrum of the corresponding difference
operator turns out to be discrete (instead of continuous like the
corresponding differential operator) owing to the behavior of its
eigenfunctions at $v\simeq\epsilon$ (i.e. $\mu\simeq\epsilon$).
Therefore, we expect, following the results of Ref. \cite{cfrw},
that the eigenvalues $\omega$ of the difference operator $\hat{\mathfrak{h}}^{{\rm dcr}}$
belong to a countable set, which we will call $\{\omega(n)\}$. One
cloud also expect that the possible (positive) values of $\omega(n)$
will depend on $\epsilon\in(0,4\rho]$, and for a given $\epsilon$,
they will also depend on $k$ and $M$. Let us comment that the particular
values of the sequence $\{\omega(n)\}$ as well as the explicit form
of the eigenfunctions $\phi_{\omega}^{{\rm dcr}}(k,\mu,M)$, to the
knowledge of the authors, can only be determined numerically by now,
unless new analytical tools are developed. In addition, a second look
on the difference equation \eqref{eq:diff-op-ext} tell us that the
eigenfunctions are completely determined by their value at the initial
data section $\phi_{\omega}^{{\rm dcr}}(k,\mu=\epsilon,M)$. Therefore,
the spectrum of $\hat{\mathfrak{h}}^{{\rm dcr}}$ will be non-degenerated.
Moreover, let us recall that, if we choose the initial data to be real,
all the coefficients $\phi_{\omega}^{{\rm dcr}}(k,\mu,M)$ for any
$\mu$ will be also real.

Eventually, the corresponding eigenfunctions, as
functions of $\mu$, will be square summable, fulfilling the normalization
condition 
\begin{align}\nonumber
\langle\phi_{\omega_{n}}^{{\rm dcr}}|\phi_{\omega_{n'}}^{{\rm dcr}}\rangle=\sum_{{\rm n}}\phi_{\omega_{n}}^{{\rm dcr}}(k,\epsilon+4{\rm n}\rho,M)\\
\times\phi_{\omega_{n'}}^{{\rm dcr}}(k,\epsilon+4{\rm n}\rho,M)=\delta_{nn'},
\end{align}
recalling that these coefficients are real. It is worth commenting
that due to the scaling \eqref{eq:scaling-out}, the coefficients
in the previous sum are weighted simply with the unit. This is not
the case, for instance, in Ref. \citep{cfrw} where the norm of the
corresponding eigenfunctions includes a weight function different
from the unit, since no scalings of the solutions are considered. 

The constraint in this basis takes the algebraic form
\begin{equation}\label{eq:disc-const}
\omega(n)-\Delta k^{2}=0.
\end{equation}

Let us now study the solutions to the constraint
when $k<M/2\hbar\lambda$. In this case we will
follow again the ideas of \cite{Martin-Benito2009} in order to render
our equation in a suitable representation where it will again become
separable. Let us comment that the solutions to the difference equation
for $k<M/2\hbar\lambda$ have different asymptotic
behaviors at $\mu\to\infty$ than the ones for $k> M/2\hbar\lambda$.
It involves that the change of representation that will be considered
in each case, in order to express the constraint equation in a simple
separable form, must not be the same.

With all this in mind, let us consider this invertible
scaling 
\begin{equation}
\phi_{j}^{{\rm cnt}}(k,\mu,M)=(\hbar G_2\hat{\mu})^{1/2}\phi(k,\mu,M).\label{eq:scaling-mu}
\end{equation}
As before, $\mu=0$ could be problematic in order to define this redefinition
properly. But let us recall that this sector has been decoupled from
the quantum theory. In consequence $\mu$ has a global minimum equal
to $\epsilon>0$. Therefore, the previous scaling \eqref{eq:scaling-mu}
can be inverted and the original description recovered. The new functions
$\phi^{{\rm cnt}}(k,\mu,M)$ now fulfill the difference equation 
\begin{widetext}
\begin{equation}
-f_{+}^{{\rm cnt}}(\mu-4\rho)\phi^{{\rm cnt}}(k,\mu-4\rho,M)-f_{-}^{{\rm cnt}}(\mu+4\rho)\phi^{{\rm cnt}}(k,\mu+4\rho,M)^{{\rm cnt}}+f_{0}^{{\rm cnt}}(k,\mu,M)\phi^{{\rm cnt}}(k,\mu,M)=0.\label{eq:2vs-dcr-diffc}
\end{equation}
but this time the coefficients are 
\begin{align}
f_{\pm}^{{\rm cnt}}(\mu)= & \frac{1}{16\rho^{2}}|\mu|^{-1/4}|\mu\pm2\rho|^{1/2}|\mu\pm4\rho|^{-1/4}\left[\sgn(\mu\pm4\rho)+\sgn(\mu\pm2\rho)\right]\left[\sgn(\mu\pm2\rho)+\sgn(\mu)\right],\label{eq:f+-densitized-2}\\
f_{0}^{{\rm cnt}}(\mu,k,M) & =\frac{1}{16\rho^{2}\mu}\left[(|\mu||\mu+2\rho|)^{1/2}\left[\sgn(\mu)+\sgn(\mu+2\rho)\right]^{2}\right.\nonumber \\
& \left.+(|\mu||\mu-2\rho|)^{1/2}\left[\sgn(\mu)+\sgn(\mu-2\rho)\right]^{2}\right]-\frac{{\rm sgn}(\mu)}{\mu\rho^{2}}\Delta k^{2}(|\mu+\rho|^{1/2}-|\mu-\rho|^{1/2})^{2},
\end{align}
\end{widetext}
This version of the scalar constraint, as we mentioned previously,
can be naively understood as a densitized version of the original classical
constraint after the choice of $N\mu^{-1}$ as the new lapse function
(and an adequate factor ordering and a global factor $\hbar G_2$). Following the notation that we
introduced, we will denote this new scalar constraint by $\hat{\mathfrak{H}}^{{\rm cnt}}$.
It is related with the original one by means of 
\begin{equation}
\hat{\mathfrak{H}}^{{\rm cnt}}=(\hbar G_2\hat{\mu})^{-1/2}\hat{\mathfrak{H}}(\hbar G_2\hat{\mu})^{-1/2}.
\end{equation}
The difference operator that will be studied now reads 
\begin{equation}
\hat{\mathfrak{h}}^{{\rm cnt}}=\hat{\mathfrak{H}}^{{\rm cnt}}-\left(1-\frac{G_2M}{2\hbar G_2\lambda k}\right)(\hbar G_2\lambda)^{2}k^{2}.\label{eq:in-eigen-eq}
\end{equation}
Therefore, we have written again the original constraint $\hat{\mathfrak{H}}$
into a suitable separable form according to the condition $k<M/2\hbar\lambda$.

We will now study the spectrum of the difference
operator $\hat{\mathfrak{h}}^{{\rm cnt}}$, by means of eigenvalue
problem 
\begin{equation}
\hat{\mathfrak{h}}^{{\rm cnt}}|\phi_{\omega}^{{\rm cnt}}\rangle=\omega|\phi_{\omega}^{{\rm cnt}}\rangle,\label{eq:int-diff-op}
\end{equation}
for $\omega\geq0$, which are the physically interesting values. This
equation can be written in the form of \eqref{eq:2vs-dcr-diffc},
but with coefficients
\begin{widetext}
\begin{align} \nonumber
&\tilde{f}_{\pm}^{{\rm cnt}}(\mu)=  \frac{1}{16\rho^{2}}|\mu|^{-1/4}|\mu\pm2\rho|^{1/2}|\mu\pm4\rho|^{-1/4}\left[\sgn(\mu\pm4\rho)+\sgn(\mu\pm2\rho)\right]\\ 
&\times\left[\sgn(\mu\pm2\rho)+\sgn(\mu)\right],\\
&\tilde{f}_{0}^{{\rm cnt}}(\mu,k,M,\omega)  =\frac{1}{16\rho^{2}\mu}\left[(|\mu||\mu+2\rho|)^{1/2}\left[\sgn(\mu)+\sgn(\mu+2\rho)\right]^{2}\right.\nonumber \\
& \left.+(|\mu||\mu-2\rho|)^{1/2}\left[\sgn(\mu)+\sgn(\mu-2\rho)\right]^{2}\right]-\frac{{\rm sgn}(\mu)}{|\mu|\rho^{2}}\Delta k^{2}(|\mu+\rho|^{1/2}-|\mu-\rho|^{1/2})^{2}-\omega,
\end{align}
\end{widetext}
Let us recall, again, that the coefficients $\phi_{\omega}^{{\rm cnt}}(k,\mu,M)$
of these eigenstates are determined by their initial data $\phi_{\omega}^{{\rm cnt}}(k,\mu=\epsilon,M)$
through the difference equation \eqref{eq:int-diff-op}. In consequence,
the (positive) spectrum of $\hat{\mathfrak{h}}^{{\rm cnt}}$ will
be non-degenerated. Moreover, the coefficients $\phi_{\omega}^{{\rm cnt}}(k,\mu,M)$
will be real if $\phi_{\omega}^{{\rm cnt}}(k,\mu=\epsilon,M)\in\mathbb{R}$,
since the previous functions $\tilde{f}_{0}^{{\rm cnt}}$ and $\tilde{f}_{\pm}^{{\rm cnt}}$
are also real.

We will assume, again, that these solutions have
a well defined and smooth asymptotic behavior for $\mu\to\infty$. Let us recall that this involves eigenvalues with typical scales much bigger than $4\rho$. This continuity condition allows us to approximate the difference equation for those large scale eigenvalues at $\mu\to\infty$
by 
\begin{widetext}
\begin{align}
 0=&-\tilde{f}_{+}^{{\rm cnt}}(\mu-4\rho)\phi_{\omega}^{{\rm cnt}}(k,\mu-4\rho,M)-\tilde{f}_{-}^{{\rm cnt}}(\mu+4\rho)\phi_{\omega}^{{\rm cnt}}(k,\mu+4\rho,M)+\tilde{f}_{0}^{{\rm cnt}}(k,\mu,M,\omega)\phi_{\omega}^{{\rm cnt}}(k,\mu,M)\nonumber \\
=& -4\partial_{\mu}^{2}\phi_{\omega}^{{\rm cnt}}(k,\mu,M)-\frac{\gamma^{2}}{\mu^{2}}\phi_{\omega}^{{\rm cnt}}(k,\mu,M)-\omega\phi_{\omega}^{{\rm cnt}}(k,\mu,M)+{\cal O}(\rho^{2}/\mu^{2}),\label{eq:bessel-diff-op}
\end{align}
\end{widetext}
but this time with $\gamma^{2}=\Delta k^{2}+3/4$. Let us comment
that this very same differential equation would have been obtained
if we would have considered a WDW representation, instead of the loop
quantization, with a suitable choice in the ordering of the operators
for the definition the corresponding Hamiltonian constraint.

The solutions to this differential equation are linear
combinations of Hankel functions of first $H_{i\gamma}^{(1)}(y)$
and second $H_{i\gamma}^{(2)}(y)$ kind, multiplied by a factor $y^{1/2}$,
where $y=\mu\sqrt{\omega}/2$. In consequence, the asymptotic limit
of the eigenstates will be 
\begin{align}\nonumber
\lim_{\mu\to\infty}\phi_{\omega}^{{\rm cnt}}(k,\mu,M)=&Ay^{1/2}H_{i\gamma}^{(1)}(y)\\
&+By^{1/2}H_{i\gamma}^{(2)}(y).\label{eq:cnt-asymp-lim}
\end{align}
These functions have a well known asymptotic limit at $y\to\infty$,
corresponding to 
\begin{equation}
H_{i\gamma}^{(1)}(y)=\sqrt{\frac{2}{\pi y}}e^{i(y-\pi/4+\gamma\pi/2)},
\end{equation}
with $H_{i\gamma}^{(2)}(y)=\big(H_{i\gamma}^{(1)}(y)\big)^{*}$. This asymptotic limit of the Hankel functions, together with \eqref{eq:cnt-asymp-lim}
and the fact that $\phi^{{\rm cnt}}(k,\mu,M)\in\mathbb{R}$ at any
$\mu$, allow us to conclude that 
\begin{equation}
\lim_{\mu\to\infty}\phi_{\omega}^{{\rm cnt}}(k,\mu,M)=A\cos\left[\frac{\sqrt{\omega}}{2}\mu+\beta\right],\label{eq:eigen-int-prescB}
\end{equation}
with $A$ a normalization constant and $\beta$ a phase that it is
expected to depend on $\Delta k$, and $\epsilon$. This asymptotic
behavior is radically different in comparison with the eigenstates $\phi_{\omega}^{{\rm dcr}}(k,\mu,M)$.
Instead of decaying exponentially, they simply oscillate as standing
waves (up to negligible corrections) of frequency $\sqrt{\omega}/2$.
Therefore, our experience in loop quantum cosmology \cite{aps-old,aps-imp,Martin-Benito2009}
tell us that these eigenfunctions will be normalizable functions of
$\mu$ (in the generalized sense) 
\begin{align}\nonumber
&\langle\phi_{\omega}^{{\rm in}}|\phi_{\omega'}^{{\rm in}}\rangle=\sum_{{\rm n}}\phi_{\omega}^{{\rm cnt}}(k,\epsilon+4{\rm n}\rho,M)\\
&\times\phi_{\omega'}^{{\rm cnt}}(k,\epsilon+4{\rm n}\rho,M)=\delta\left(\sqrt{\omega}/2-\sqrt{\omega'}/2\right).
\end{align}

Eventually, the constraint in the basis of states $|\phi_{\omega}^{{\rm in}}\rangle$ takes the form
\begin{equation}
\omega+\left(1-\frac{G_2M}{2\hbar G_2\lambda k}\right)(\hbar G_2\lambda)^{2}k^{2}=0.\label{eq:cont-const}
\end{equation}

It is worth commenting that these results might be modified for those ``high frequency'' eigenvalues, where the discreteness of the lattice in $\mu$ is important. This will be a matter of future research.

\bibliography{main}

\end{document}